\documentclass[9pt,lineno]{style}

\usepackage{lipsum} 
\usepackage[version=4]{mhchem}
\usepackage{siunitx}
\usepackage{bm}
\usepackage{hyperref}
\usepackage{float}
\usepackage{amsmath,mathtools}

\usepackage{setspace}

\renewcommand{\L}{\mathcal{L}}
\newcommand{\T}{\mathcal{T}}

\graphicspath{ {./fig3/} }

\DeclareSIUnit\Molar{M}

\DeclareUnicodeCharacter{2212}{-}

\newcommand{\remove}[1]{}

\title{
Energy budgets govern synaptic precision and its regulation during plasticity}
\usepackage{parskip}

\author[1*]{James Malkin}
\author[2]{Cian O'Donnell}
\author[1]{Conor Houghton}
\affil[1]{Faculty of Engineering, University of Bristol, Bristol, UK}
\affil[2]{Intelligent Systems Research Centre, School of Computing, Engineering, and Intelligent Systems, Ulster University, Derry/Londonderry, UK}

\corr{james.malkin@bristol.ac.uk}{JM}

\begin{document}
\maketitle
\begin{abstract}

Synaptic transmission must balance the need for reliable signalling against the metabolic cost of achieving that reliability. How energetic constraints shape synaptic precision and its regulation during plasticity remains unclear. Here we develop an energy-constrained framework in which synapses minimise postsynaptic response variance subject to a fixed mean and an effective energy budget. Combinations of candidate physiological costs are used to estimate an energy cost for synaptic transmission; this cost is then inferred from quantal statistics. Analysing five published pre- and post-plasticity datasets, we find that observed synaptic mean–variance pairs cluster near a minimal-energy boundary, indicating that precision is limited by energetic availability. Model comparison identifies a dominant calcium pump–like cost paired with a smaller vesicle turnover–like cost, yielding a separable precision–energy relationship, $\sigma^{-2}\propto E^5$. We further show that plasticity systematically updates synaptic energy budgets according to the scale-free magnitude of mean change, enabling accurate prediction of post-plasticity variance from energy allocation alone. These results provide direct experimental support for the hypothesis that synaptic precision is governed by energy budgets, establishing energy allocation as a fundamental principle linking metabolic constraints, synaptic reliability, and plasticity.
\end{abstract}

%%%%

\section{Introduction}

Synapses are intrinsically noisy---vesicle release is probabilistic, vesicle pools are finite, and postsynaptic responses fluctuate across trials even when pre- and postsynaptic activity are repeated \citep{larkman1992presynaptic, allen1994evaluation, huang1997estimating, van2003effects}. If this variability is too large, transmission becomes unreliable and information processing degrades \citep{zador1996vc,zador1998impact, nolte2019cortical}; if it is too small, achieving and maintaining such precision may incur substantial metabolic cost \citep{levy2002energy,harris2015energy,harris2019energy,malkin2024signatures}. Synapses must therefore regulate reliability under energetic constraint. How this regulation emerges, and whether synaptic precision reflects an underlying energetic principle, remains unclear.

A classical and experimentally grounded description of synaptic transmission is provided by the quantal model \citep{katz1972statistical}, in which synaptic responses arise from the stochastic release of discrete neurotransmitter packets. In this framework, synaptic efficacy and variability are determined by three quantal parameters: the number of release-ready vesicles or release sites ($n$), the release probability ($p$), and the quantal size ($q$), which together determine the mean and variance of postsynaptic responses \citep{del1954quantal,bliss1993synaptic,scheuss2002separation,brock2020practical}. These parameters reflect identifiable biological processes, including vesicle availability, presynaptic release machinery, and postsynaptic receptor function \citep{malinow2002ampa,korber2016molecular}. Quantal statistics therefore provide a mechanistically interpretable description of synaptic transmission and an experimentally accessible link between physiological state and functional reliability.

Importantly, quantal parameters do not vary independently. Across diverse preparations, empirical studies show systematic co-variation between $n$, $p$, and $q$ during plasticity and across synapses \citep{costa2015unified,costa2017synaptic,biederer2017transcellular,gou2022re}. This coordinated variation suggests that synaptic reliability is regulated subject to underlying constraints, rather than adjusted arbitrarily through independent molecular mechanisms. However, existing theories primarily describe how synaptic mean strength changes during plasticity \citep{malenka2004ltp,abbott2004synaptic}, and provide limited explanation for how synaptic variability is regulated or why synapses exhibit particular reliability levels.

Several lines of evidence suggest that reliability is constrained by energetic considerations. Increasing release probability requires elevated presynaptic Ca$^{2+}$ influx and ATP-dependent clearance; maintaining vesicle pools and release readiness requires continual vesicle cycling, membrane maintenance, and protein turnover. These processes impose substantial energetic costs \citet{attwell2001energy, engl2015non,karbowski2019metabolic}). Consistent with this view, theoretical work has shown that optimising reliability under energetic constraint produces heterogeneous synaptic noise levels consistent with Bayesian learning principles \citep{malkin2024signatures,rusakov2020noisy,aitchison2021synaptic}. These results suggest that synaptic variability may reflect an optimal allocation of metabolic resources, rather than unavoidable biological imprecision.

Here we test the hypothesis that synaptic precision is constrained by an effective energy budget that limits attainable reliability. We propose that each synapse operates near a minimum-energy boundary determined by its mean strength and available metabolic resources. In this framework, energy is treated as a latent state variable inferred from experimentally measured synaptic statistics. The quantal parameters $(n,p,q)$ provide a physiological basis for constructing candidate energy cost models, enabling quantitative inference of effective synaptic energy budgets from observed mean and variance.

A key prediction of this framework is that synaptic precision obeys a strict relationship with energy availability: at fixed mean strength, increasing precision requires increasing energetic investment. Furthermore, if plasticity reallocates metabolic resources, changes in synaptic strength should be accompanied by predictable changes in energy budget and, consequently, synaptic variability.

In this study, we develop an energy-constrained framework that combines biophysically motivated cost models with experimental measurements of synaptic transmission. Using five published datasets reporting pre- and post-plasticity synaptic mean and variance, we infer effective synaptic energy budgets and test whether observed synapses operate near the predicted minimum-energy boundary. We then determine how synaptic energy budgets change during plasticity and test whether these budget changes predict post-plasticity variability. This approach enables us to identify physiologically consistent energy cost structure, derive quantitative relationships linking energy and synaptic precision, and test whether synaptic noise can be predicted from energy allocation alone.

Together, these results provide direct experimental evidence that synaptic precision is constrained and regulated by energy availability. By establishing energy budgets as a quantitative predictor of synaptic variability, this work identifies energy allocation as a fundamental organising principle linking metabolic constraint, synaptic reliability, and plasticity.

\section{Biophysical origin of energy costs}
\label{sec:biophysical_costs}

Our framework requires a mapping from synaptic state to metabolic expenditure.
Rather than assuming a single ground-truth energy function, we specify a small set of
candidate cost components motivated by biophysical scaling arguments, and later
identify which combinations best explain observed quantal configurations under the
minimum-energy boundary hypothesis (also see \hyperref[methods_energy_cost]{Methods -- Energy cost selection and rational}).
We treat $E$ as an effective energy budget in arbitrary units, capturing relative energetic pressure.
We model synaptic energy as a mixture of component costs, $C_i$,
\begin{equation}
E(n,p,q)=\sum_{i}\beta_i\,C_i(n,p,q),
\qquad \beta_i\ge 0,\qquad \sum_i \beta_i = 1,
\label{eq:energy_sum}
\end{equation}
where each $C_i$ corresponds to an identifiable process (e.g.\ calcium handling, turnover),
and the weights $\beta_i$ encode their relative contribution in the effective budget.
The role of this section is to define the candidate components and justify their functional
dependence on $(n,p,q)$.

\subsection{Quantal parameters and experimentally accessible statistics}
\label{subsec:quantal_model}
We describe evoked synaptic responses using a classical quantal model in which a presynaptic spike
triggers probabilistic release of neurotransmitter packets.
In a binomial approximation a synapse has $n$ release-ready sites, each releases with probability $p$,
and each successful release produces a postsynaptic response of mean amplitude $q$.
Under standard assumptions,
\begin{equation}
\mu = npq,
\qquad
\sigma^2 = np(1-p)q^2,
\label{eq:quantal}
\end{equation}
where $\mu$ and $\sigma^2$ denote the mean and variance of the evoked PSP.
Although real synapses violate strict binomial assumptions (heterogeneity across release sites, multivesicular
release, nonlinear summation \citep{quastel1997binomial, cushing2005effect, rudolph2015ubiquitous}), quantal decompositions remain empirically useful for linking
changes in mean/variance to mechanistic expression in $(n,p,q)$ \citep{herreras1987characteristics,faber1991applicability,enoki2009expression}.

In our framework the central latent variable is the energy budget $E$,
which constrains feasible quantal states and therefore limits achievable reliability at a given mean.
The candidate costs below define alternative hypotheses for how energetic pressure penalises different
reliability strategies.

\subsection{Candidate physiological contributors}
Synapses incur energetic costs through both activity-dependent signalling and
activity-independent maintenance. We focus on costs that plausibly trade off against reliability
for evoked transmission and admit interpretable scaling with $(n,p,q)$, identified in \citet{malkin2024signatures}.
We consider five canonical contributors: (i) presynaptic calcium pumping associated with release
probability; (ii) vesicle membrane maintenance (including reversal of proton leak);
(iii) actin/cytoskeletal scaffolding supporting vesicle pools; (iv) vesicle trafficking/recruitment;
and (v) protein turnover maintaining release machinery and vesicle readiness.
These components are not intended as a complete molecular accounting (for this, see \citet{attwell2001energy, engl2015non,karbowski2019metabolic}), but as a parsimonious basis set
with distinct predicted reliability penalties that can be compared empirically.

\begin{table}[H]
\centering
\begin{tabular}{@{}llll@{}}
\toprule
Process & Symbol & Cost component $C_i(n,p,q)$ & Key dependence \\ \midrule
Ca$^{2+}$ pumping (release support) & $C_p$ & $\left[p/(1-p)\right]^{1/4}$ & steep in $p$ \\
Vesicle membrane / proton leak & $C_m$ & $n\,q^{2/3}$ & surface-area scaling \\
Actin/cytoskeletal scaffolding & $C_a$ & $n\,q^{1/3}$ & linear in radius \\
Trafficking / mobilisation & $C_{tr}$ & $np$ & released-vesicle count \\
Protein turnover / readiness & $C_t$ & $n$ & vesicle-number scaling \\ \bottomrule
\end{tabular}
\caption{Candidate biophysical cost components used to construct the effective synaptic energy budget in
Eq.~\eqref{eq:energy_sum}. Each component is motivated by a scaling argument linking expenditure to quantal
state. See \hyperref[subsec:cost_components]{Methods -- Candidate cost components and dependence on $(n,p,q)$}.}
\label{tab:costs}
\end{table}

\subsection{From cost components to reliability penalties}
\label{subsec:reliability_penalties}
On the fixed-mean manifold $\mu=npq=\mu^\ast$, reducing variance typically requires increasing $np$ and
adjusting $q$ accordingly. Each cost component therefore induces a distinct penalty for increasing
precision. Increasing $p$ reduces $\sigma^2$ but increases $C_p$ sharply; increasing $n$ reduces $\sigma^2$
but increases $n$-dependent maintenance terms; and changing $q$ trades amplitude against membrane/scaffold
costs. These induced penalties are the mechanism by which metabolic pressure constrains plasticity
expression: different costs imply different energy--precision frontiers and different preferred directions
in $(n,p,q)$ space.

\section{Results}

\subsection{The plasticity data}
\label{plasticity_data}

\begin{figure}[H]
\centering
\includegraphics[width=1.0\textwidth]{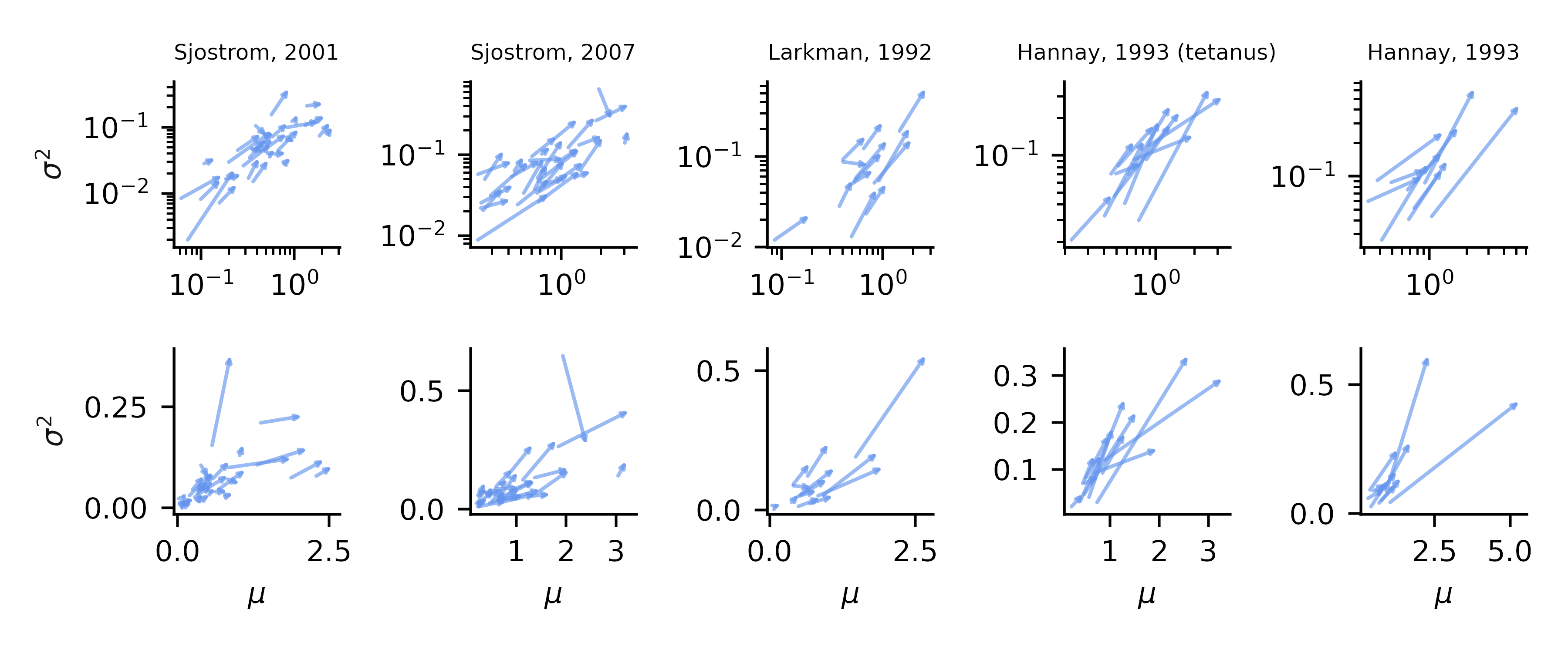}
\caption[Plasticity-induced changes in synaptic mean and variance]{\textbf{Plasticity-induced changes in
synaptic mean and variance.} Each arrow shows the transition for a single synapse from baseline
$(\mu_0,\sigma_0^2)$ to post-plasticity $(\mu_1,\sigma_1^2)$ across five datasets. Top row: log--log
coordinates; bottom row: linear coordinates.}
\label{fig_mu_var_true}
\end{figure}

To test whether energetic constraints can account for \emph{plasticity-induced} changes in synaptic
variability and quantal expression, we analysed five published datasets that report baseline and
post-induction synaptic statistics for individual synapses. For each synapse we extracted the paired
transition
$(\mu_0,\sigma_0^2)\;\rightarrow\;(\mu_1,\sigma_1^2)$
where $\mu$ is mean synaptic efficacy and $\sigma^2$ is trial-to-trial variance of the evoked response
(see \hyperref[subsec:cost_components]{Methods -- The plasticity data} for extraction details).
Figure.~\ref{fig_mu_var_true} summarises these transitions: each arrow depicts one synapse, mapping its
baseline state to its post-plasticity state (top: log--log; bottom: linear).

\paragraph{Datasets.}
The five datasets span distinct preparations and induction protocols, including spike-timing-dependent
plasticity (STDP), LTP, and short-lived potentiation. We summarise them briefly here and provide full
protocol details in Materials and Methods.

\vspace{4pt}
\noindent\textbf{Sj\"ostrom et al.\ (2001).}
STDP was induced in layer~5 pyramidal neurons in visual cortex slices by paired pre/post spiking evoked via
current injection across multiple pairing frequencies and timing offsets. EPSP amplitudes were recorded
with whole-cell patch clamp \citep{sjostrom2001rate}.

\vspace{4pt}
\noindent\textbf{Sj\"ostrom et al.\ (2007).}
Whole-cell EPSP recordings were obtained from layer~5 neurons in visual cortex slices. Potentiation was
induced by prolonged pairing protocols with varying stimulation duration and amplitude
\citep{sjostrom2007multiple}.

\vspace{4pt}
\noindent\textbf{Larkman et al.\ (1992).}
LTP was induced in CA1 hippocampus using tetanic presynaptic stimulation paired with simultaneous tetanus
of additional pyramidal fibres. EPSPs were measured with sharp-electrode recordings
\citep{larkman1992presynaptic}.

\vspace{4pt}
\noindent\textbf{Hannay et al.\ (1993; tetanus).}
Short-lived potentiation was induced in CA1 hippocampus via electrode-driven tetanic stimulation. EPSPs
were measured with sharp-electrode recordings and decayed back toward baseline within minutes. The authors
note consistency with post-tetanic potentiation (PTP), compatible with residual presynaptic Ca$^{2+}$
accumulation during tetanus \citep{hannay1993common}.

\vspace{4pt}
\noindent\textbf{Hannay et al.\ (1993; current injection).}
Short-lived potentiation was induced while maintaining baseline presynaptic stimulation and depolarising
the postsynaptic neuron with sustained ($\sim$40\,s) current injection. EPSPs were measured with
sharp-electrode recordings and decayed toward baseline within minutes \citep{hannay1993common}.

In the results that follow we treat $(\mu_0,\sigma_0^2)$ and $\mu_1$ as observed inputs, infer an effective
baseline budget $E_0$ from $(\mu_0,\sigma_0^2)$ under a candidate cost model, predict the post-event budget
$E_1$ from the magnitude of the mean change, and then predict post-plasticity variability and quantal
expression by solving the energy-constrained noise minimisation problem at fixed $(\mu_1,E_1)$.
Critically, $\sigma_1^2$ is reserved for evaluation.

\subsection{Reliability--efficiency tradeoff}
\label{tradeoff_section}

Synaptic transmission is intrinsically stochastic, yet reliable signalling is essential for stable
computation and learning. Increasing precision (reducing trial-to-trial variability) is possible by
adjusting quantal parameters---for example by increasing vesicle availability or release probability---but
such adjustments incur metabolic costs. We formalise this tension by
introducing a synapse-local objective that trades off reliability against energetic expenditure. Throughout
this subsection we consider a single synapse in isolation, the objective depends only on that
synapse's state and does not invoke a network-level loss.

\paragraph{Tradeoff objective.}
We model the effective synaptic weight as a Gaussian random variable $w\sim\mathcal N(\mu,\sigma^2)$ and
consider a quadratic performance cost with minimiser $\hat\mu$. The reliability--efficiency objective is
\begin{align}
\mathcal T
&= \underbrace{\Big\langle (w-\hat\mu)^2 \Big\rangle}_{\text{performance}} \;+\; \gamma\,\underbrace{E}_{\text{energy}}
\label{eq:T_def}
\\[-2pt]
&= \sigma^2 + (\mu-\hat\mu)^2 + \gamma E,
\label{eq:T_expand}
\end{align}
where $\langle\cdot\rangle$ denotes expectation and the expansion in Eq.~\eqref{eq:T_expand} follows from
$\langle (w-\hat\mu)^2\rangle=\mathrm{Var}(w)+(\mathbb E[w]-\hat\mu)^2$.
The scalar $\gamma>0$ sets the exchange rate between units of variance and units of energy.
Energy is written as the mixture of candidate costs introduced in \hyperref[sec:biophysical_costs]{Biophysical origin of energy costs}, Eq.~\eqref{eq:energy_sum}.
% \begin{align}
% E=\sum_i \beta_i C_i(n,p,q), \qquad \beta_i\ge 0,\quad \sum_i\beta_i=1.
% \label{eq:E_def}
% \end{align}

\paragraph{Fixing the mean to isolate reliability.}
The objective Eq.~\eqref{eq:T_expand} trades off performance and energy, and its minimiser with respect to the
mean,
\(
\mu^*=\arg\min_\mu \mathcal T,
\)
generally differs from the purely performance-optimal value $\hat\mu$ because the energetic cost depends on
$\mu$.
We interpret $\mu^*$ as the mean efficacy that optimally balances accuracy and energetic expenditure.

In our analysis, we assume that the empirical post-plasticity mean approximates this tradeoff optimum,
$\mu_{\mathrm{emp}}\approx\mu^*$, reflecting plasticity under energetic constraint.
Conditioning on the observed mean therefore amounts to conditioning on the solution of the full
performance--energy optimisation problem.

Operationally, this yields a separation of roles: plasticity sets the mean efficacy through optimisation of
$\mathcal T$, while the remaining degrees of freedom in $(n,p,q)$ tune reliability around this operating
point.
Under this assumption, the mean-error term $(\mu-\hat\mu)^2$ is constant, and the
objective reduces to
\begin{align}
\mathcal T = \sigma^2 + \gamma E + \mathrm{const}.
\label{eq:T_tradeoff}
\end{align}
Thus, at fixed $\mu^*$, minimising $\mathcal T$ amounts to selecting quantal parameters that reduce variance
while paying an energetic penalty.

\subsection{Robust two-term approximation: calcium pumping + turnover.}
\label{subsec:pump_turnover_choice}
The full energy model in Eq.~\eqref{eq:energy_sum} is useful for completeness and model comparison, but much of the
analytic structure is captured by a two-term approximation consisting of a calcium-dependent release cost
paired with a single $n$-dependent term,
\begin{equation}
E(n,p,q) \approx \beta_p\,C_p(p) + (1-\beta_p)\,C_n(n).
\label{eq:two_term}
\end{equation}
In \hyperref[methods_energy_cost]{Methods -- Energy cost selection and rationale} we compare alternative parameterisations of Eq.~\eqref{eq:energy_sum} systematically using a two-stage
boundary-fit procedure and select the pairing that provides the most robust predictions and the most
interpretable noise--energy mapping. By using experimental synaptic measurements to infer which cost components dominate (i.e.\ to select $\beta_i$) and to test whether observed synaptic states are consistent with operating near the minimal-energy boundary implied by the model we show that the calcium pump$+$turnover form,
\begin{equation}
E(n,p,q)=\beta_p\left(\frac{p}{1-p}\right)^{1/4} + (1-\beta_p)\,n,
\label{eq:pump_turnover}
\end{equation}
is uniquely well-posed for our pipeline, yielding a separable precision--energy relation and robust
inference of post-plasticity variability and quantal expression.

This minimal two-term energy model that captures the dominant qualitative structure includes a steep cost for
increasing release probability (presynaptic Ca$^{2+}$ pumping) paired with an $n$-dependent maintenance cost
(vesicle turnover),
corresponding to Eqs.~\eqref{eq:cp} and \eqref{eq:ct}.
Using the binomial quantal model (Eq.~\eqref{eq:quantal}), fixed mean $\mu^*=npq$, and the odds
parametrisation $b\equiv \tfrac{p}{1-p}$, the normalised objective becomes
\begin{align}
\mathcal T
\;\propto\;
\underbrace{\frac{1-p}{np}}_{\sigma^2/\mu^{*2}}
\;+\;
\frac{\gamma}{(\mu^*)^2}\,
\underbrace{\Big(\beta_p\,b^{1/4} + (1-\beta_p)\,n\Big)}_{E(n,p)}
\;+\;\mathrm{const}.
\label{eq:T_np_form}
\end{align}
This expression makes the competing gradients explicit. Increasing $n$ or $p$ reduces the noise term
$\frac{1-p}{np}$, but increases energy through vesicle turnover and calcium pumping, respectively.

\begin{figure}[H]
  \centering
  \includegraphics{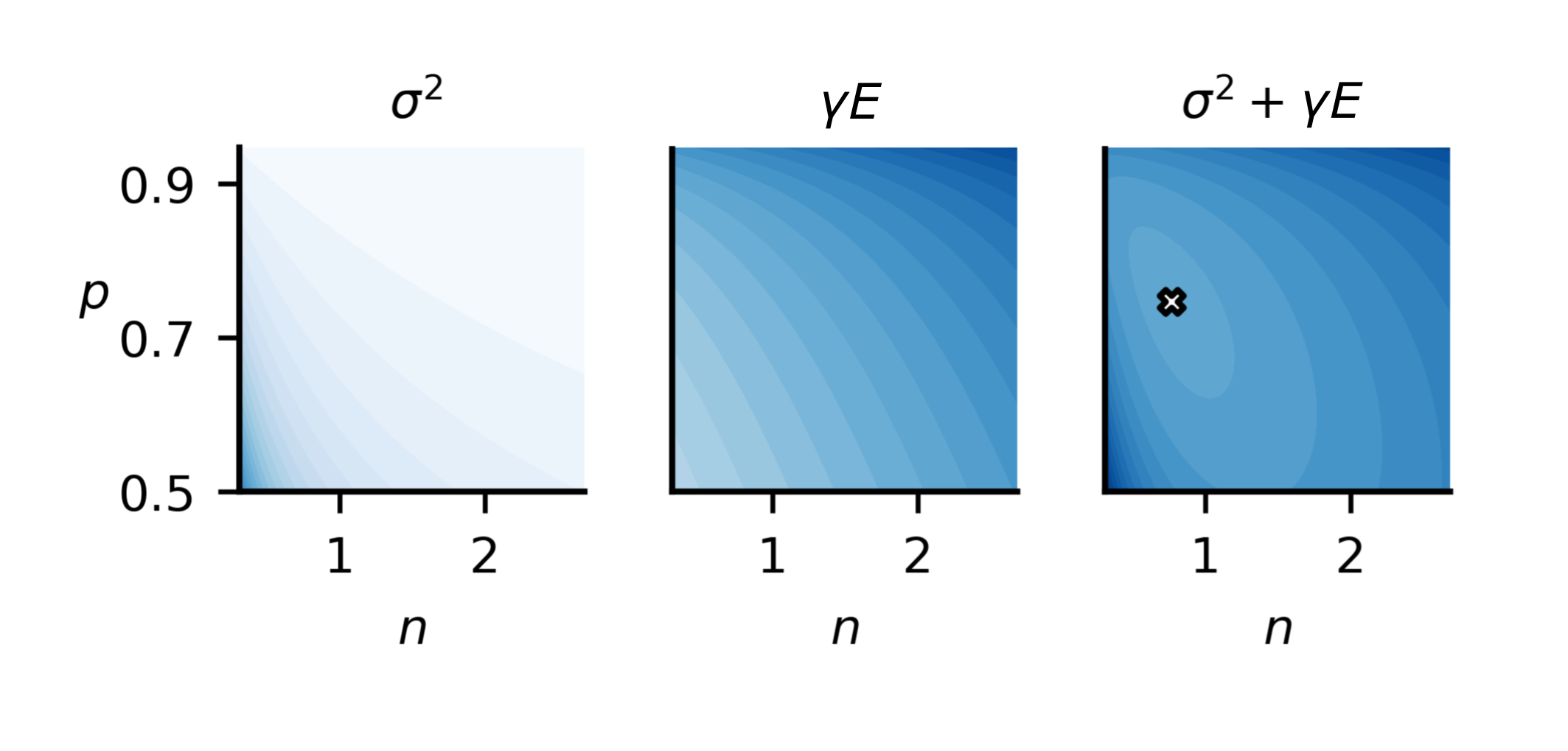}
  \caption[Tradeoff between noise cost and energy cost for synaptic plasticity]{\textbf{Tradeoff between noise cost and energy cost for synaptic plasticity.}
  Heat maps of (i) the noise term $\sigma^2$ (normalised by $\mu^{*2}$), (ii) the energy term $E$, and (iii)
  the combined objective over $(n,p)$ for the calcium pump$+$turnover model. Darker shades indicate higher
  values. The black cross marks the optimum for $\gamma=0.25$, $\mu^*=0.5$, and $\beta_p=0.7$.}
  \label{fig_loss}
\end{figure}

Figure~\ref{fig_loss} visualises this interaction. Within the plotted range, the joint objective exhibits a
single well-defined minimum (black cross) (see \hyperref[fig_loss_check]{Appendix -- Convexity conditions of the tradeoff}), illustrating that the combined noise and energy terms yield an
interior optimum rather than driving $p\to 1$ or $n\to\infty$. Intuitively, the steep pump penalty prevents
$p$ from saturating, while the vesicle-number penalty prevents unbounded vesicle recruitment.

\begin{figure}[H]
  \centering
  \includegraphics{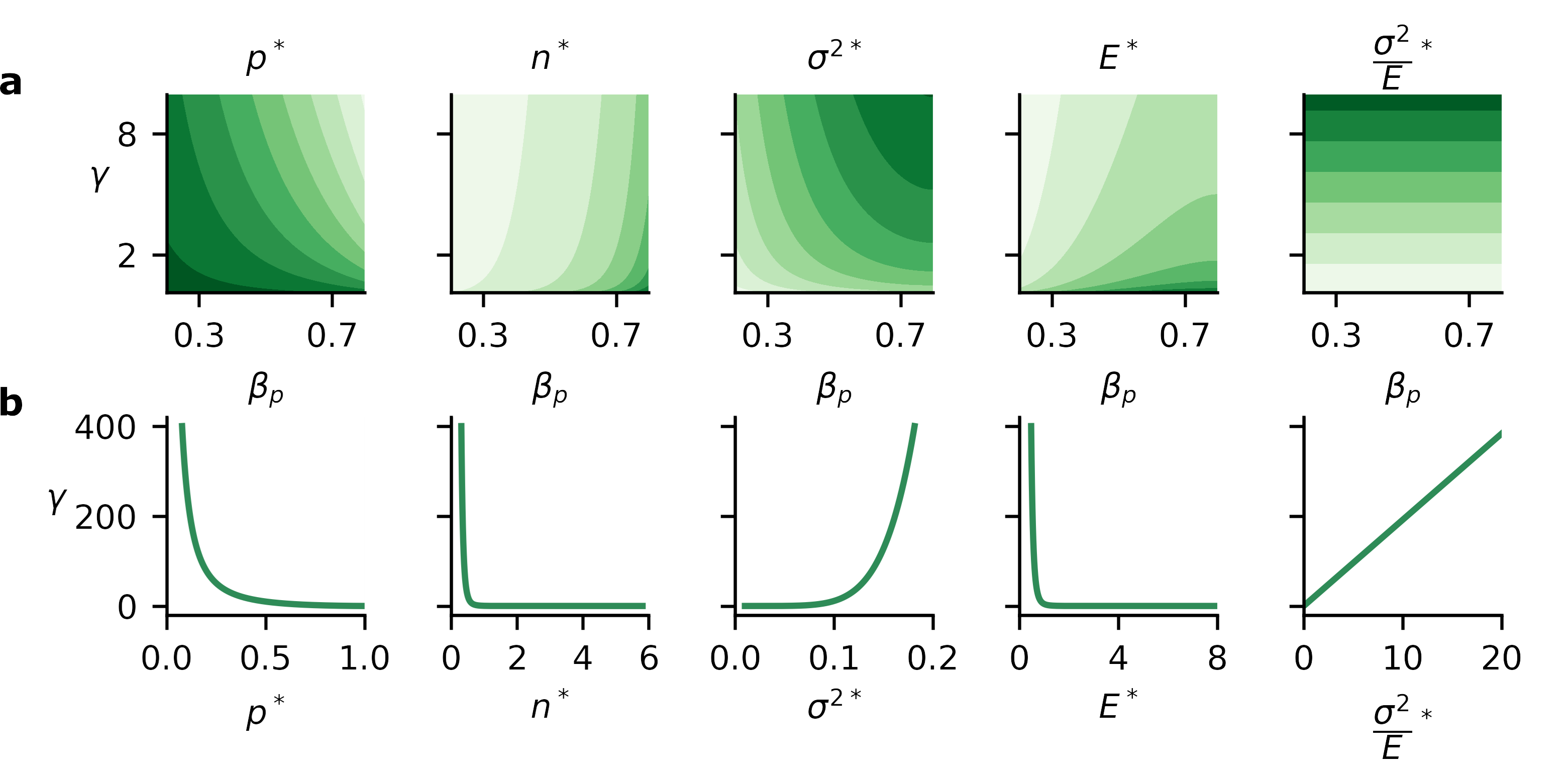}
  \caption[Impact of energy cost multiplier and scale factors on synapse expression]{\textbf{Impact of energy cost multiplier and scale factors on synapse expression.}
  \textbf{a)} Heatmaps of optimal $(p^*,n^*,\sigma^{2*},E^*,\sigma^{2*}/E^*)$ across tradeoff
  parameterisations. Darker shades of green indicate higher solutions. \textbf{b)} Dependence of $p^*,n^*,\sigma^{2*},E^*$ on $\gamma$, illustrating how
  energetic pressure shapes synaptic precision and investment.}
  \label{fig_gamma_sols}
\end{figure}

For the calcium pump+turnover model the joint objective has a unique interior minimiser
(Fig.~\ref{fig_loss}) and varies smoothly with $\gamma$ (Fig.~\ref{fig_gamma_sols}).
As energetic pressure increases ($\gamma\uparrow$) the optimum shifts toward smaller
$n^*$ and $p^*$ and therefore larger $\sigma^{2*}$; relaxing energetic pressure produces
the converse.

The dependence is captured by simple monotone scalings:
\begin{align}
E^* \propto \gamma^{-1/6},
\qquad n^* \propto \gamma^{-1/6},
\qquad
\frac{p^*}{1-p^*}\propto \gamma^{-2/3},
\qquad
\sigma^{2*}\propto \gamma^{5/6},
\label{eq:gamma_scalings}
\end{align}
for fixed $(\mu^*,\beta_p)$.
Intuitively, lowering $\gamma$ corresponds to allocating more energy to suppress variability.
Closed-form expressions are most naturally stated in the equivalent budget-constrained form and are
given in \hyperref[budget_section]{Results -- Energy budgets} (with derivations in the \hyperref[key_derivations]{Appendix -- Key derivations}).

\paragraph{How $\gamma$ shapes precision and energetic investment.}
The multiplier $\gamma$ controls the relative value placed on energy conservation versus precision.
Increasing $\gamma$ raises the penalty for energy usage, shifting the optimum toward smaller $n^*$ and $p^*$
(and consequently larger $\sigma^{2*}$); decreasing $\gamma$ yields the converse and corresponds to greater
energetic investment in reliability. Figure~\ref{fig_gamma_sols} illustrates these trends across a range of
parameterisations. In a network context $\gamma$ could vary across synapses with synapse importance; here it
acts as a single-synapse control parameter.

Unlike $\gamma$, the shares $\beta_i$ are properties of the cost model and determine which expression routes
are relatively ``cheap''. In the two-term model, increasing $\beta_p$ (placing more weight on calcium
pumping costs) shifts the optimum away from high release odds and toward vesicle-based reliability, whereas
decreasing $\beta_p$ has the opposite effect. At the optimum, the balance between these routes is captured
by
\begin{align}
\frac{\beta_p}{1-\beta_p} = \frac{4n^*}{(b^*)^{1/4}},
\qquad
b=\frac{p}{1-p},
\label{eq:beta_ratio_relation}
\end{align}
which makes explicit how energetic weighting partitions investment between vesicle availability and release
probability.

\paragraph{From tradeoff weights to budgets.}
While $\gamma$ is useful conceptually as an energy--precision exchange rate,
the empirical pipeline in this paper treats energy as a \emph{constraint}.
We therefore reparameterise the optimum in terms of a synapse-local budget $E^*$ and
derive closed-form relations $n^*(E^*)$, $p^*(E^*)$, and $\sigma^{2*}(E^*)$.
This budget formulation is equivalent to the tradeoff formulation at the optimum,
with $\gamma$ implicitly determined by $(E^*,\mu^*,\beta_p)$.

\subsection{Energy budgets}
\label{budget_section}

The tradeoff formulation above penalises energetic expenditure via the multiplier $\gamma$. For empirical applications it is often more natural to instead treat energy as a constraint: a synapse has an effective budget $E^*$ and selects $(n,p,q)$ to minimise noise subject to fixed mean and fixed energy. In this section we (i) reparameterise the optimal solutions directly in terms of $E^*$ and (ii) derive the resulting power-law mapping between budget and precision. 

\begin{figure}[H]
  \centering
  \includegraphics[width=0.85\textwidth]{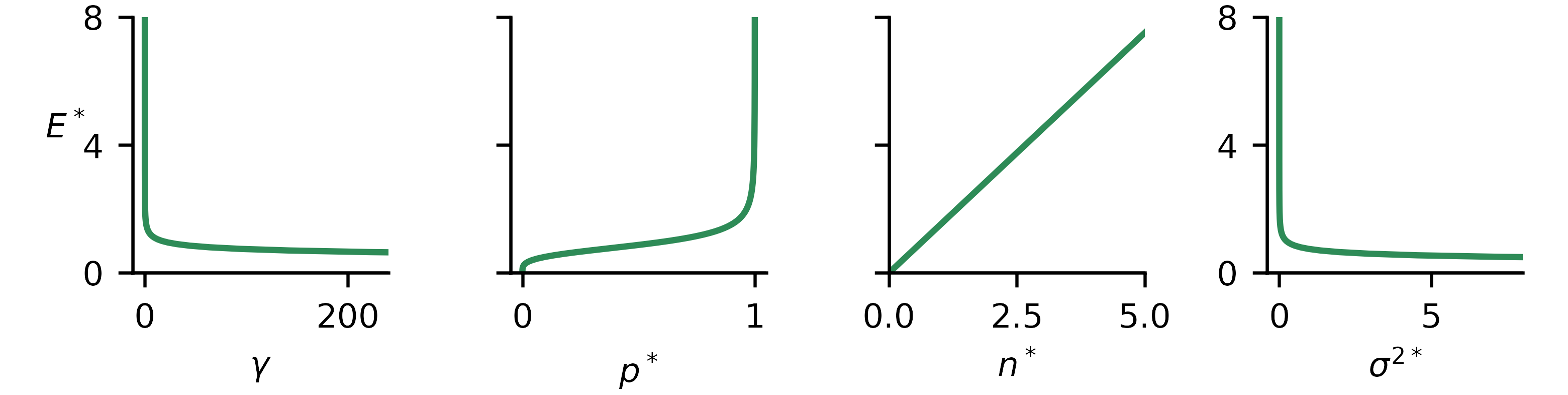}
  \caption[Energy budget-dependent predictions for quantal parameters]{\textbf{Energy budget-dependent predictions for quantal parameters.} With $\beta_p=0.95$ and $\mu^*=1$ fixed, panels show how a prescribed energy budget $E^*$ determines the optimal synaptic parameters $(n^*,p^*,\sigma^{2*})$ and the implied trade-off multiplier $\gamma$. The monotone trends reflect Eq.~\eqref{eq:proportions_rewrite}.}
  \label{fig_E_sols}
\end{figure}

\paragraph{Closed-form solutions as functions of a prescribed budget.}
We continue with the calcium pump+turnover model (Eq.~\eqref{eq:pump_turnover}) and fix $\beta_p$ (e.g.\ $\beta_p=0.95$ in Fig.~\ref{fig_E_sols}). Writing $b=\tfrac{p}{1-p}$, the optimal state under the constraints $\mu=\mu^*$ and $E=E^*$ admits closed forms (\hyperref[key_derivations]{Appendix -- Key derivations}):
\begin{align}
\label{eq:closed_forms_ET_rewrite}
E^*
  &= \Bigg[\;\frac{5^{6}}{4^{4}}\;\beta_p^{4}\,(1-\beta_p)\;\frac{(\mu^*)^{2}}{\gamma}\;\Bigg]^{\!1/6},
\\[4pt]
n^*
  &= \frac{1}{5(1-\beta_p)}\,E^*,
\\[4pt]
b^*
  &= 5(1-\beta_p)\,\kappa(\beta_p)^{-4}\,(E^*)^{4},
\\[4pt]
(\sigma^*)^{2}
  &= \kappa(\beta_p)^{5}\,\frac{(\mu^*)^{2}}{(E^*)^{5}},
\end{align}
where the (positive) budget--reliability scale factor is
\begin{align}
\kappa(\beta_p)=\frac{4}{5}\,\beta_p^{-1}\,\big(5(1-\beta_p)\big)^{4/5}>0.
\label{eq:kappa_def}
\end{align}
The first line restates, with constants, the tradeoff-optimal budget $E^*(\gamma)$ from Eq.~\eqref{eq:gamma_scalings}; the remaining lines express the optimal synaptic state directly in terms of a prescribed budget $E^*$. We display these relationships in Figure~\ref{fig_E_sols}.

These expressions make the key proportionalities transparent:
\begin{align}
(E^*)^6 \propto \frac{(\mu^*)^2}{\gamma},\qquad
n^* \propto E^*,\qquad
\frac{p^*}{1-p^*}=b^* \propto (E^*)^4,\qquad
(\sigma^*)^2 \propto \frac{(\mu^*)^2}{(E^*)^5},
\label{eq:proportions_rewrite}
\end{align}
with distinct $\beta_p$-dependent constants given explicitly by Eqs.~\eqref{eq:closed_forms_ET_rewrite}--\eqref{eq:kappa_def}. Hence, once $E^*$ and $\mu^*$ are specified, $\gamma$ is implied (via the first line) and need not be carried explicitly in the remaining predictions for $n^*$, $p^*$ and $\sigma^{2*}$.

\paragraph{Directionality of quantal changes under budget shifts.}
Tightening the budget (smaller $E^*$, equivalently larger $\gamma$) forces cheaper presynaptic configurations with fewer vesicles and lower release probability, increasing variance and requiring a compensatory increase in $q$ to maintain the same mean:
\begin{align}
E^*\downarrow\ \ (\gamma\uparrow)\quad \Rightarrow\quad
n^*\downarrow,\;\; p^*\downarrow,\;\; \sigma^{2*}\uparrow,\;\; q^*\uparrow
\qquad
\big(q^*=\mu^*/(n^*p^*)\big).
\label{eq:direction_tighten}
\end{align}
Relaxing the budget yields the opposite pattern:
\begin{align}
E^*\uparrow\ \ (\gamma\downarrow)\quad \Rightarrow\quad
n^*\uparrow,\;\; p^*\uparrow,\;\; \sigma^{2*}\downarrow,\;\; q^*\downarrow.
\label{eq:direction_relax}
\end{align}
Which variable carries most of the adjustment depends on the fixed cost weights. In the pump+turnover model, increasing the pump weight $\beta_p$ makes release probability more expensive and shifts reliability toward vesicle investment:
\begin{align}
\beta_p\uparrow\ (\beta_t\downarrow)
\quad\Rightarrow\quad
p^*\downarrow\ \ (\text{i.e.\ } b^*\downarrow),\ \ \text{typically compensated by}\ \ n^*\uparrow,
\label{eq:beta_direction}
\end{align}
consistent with the optimality relation $\frac{\beta_p}{1-\beta_p}=\frac{4n^*}{(b^*)^{1/4}}$ (Eq.~\eqref{eq:beta_ratio_relation}).

\paragraph{A budget--precision power law.}
Increasing the budget increases both vesicle number and release odds while reducing variance. The reliability scaling is particularly simple when expressed in terms of signal-to-noise $k^{-2}\equiv \mu^{2}/\sigma^{2}$:
\begin{align}
\label{eq:power_law_rewrite}
k^{-2}
= \frac{\mu^{2}}{\sigma^{2}}
= n^*\,b^*
= \Big(\tfrac{E^*}{\kappa(\beta_p)}\Big)^{5}
\;\propto\;(E^*)^{5}.
\end{align}
Thus, for fixed $\mu^*$, budgets map to precision by a strict power law: multiplicative changes in $E^*$ induce predictable multiplicative changes in $\sigma^{-2}$.

\paragraph{Log-linear form and predictable gains.}
Taking logs of Eq.~\eqref{eq:power_law_rewrite} gives
\begin{align}
\log k^{-2}=5\log E^* - 5\log \kappa(\beta_p),
\label{eq:log_power_law}
\end{align}
so $\log k^{-2}$ is affine in $\log E^*$ with slope $5$. Equivalently, for fixed $\mu^*$,
\begin{align}
\log \sigma^{-2}(hE^{*}) = \log \sigma^{-2}(E^{*}) + 5\log h.
\label{eq:log_shift}
\end{align}
Thus, the precision gain from scaling energy is uniform; doubling the budget ($h=2$) increases $\sigma^{-2}$ by a factor $2^{5}$.

\subsubsection{Summary}
The budget parameterisation yields three central results. \textit{First}, for fixed $\mu^*$ and cost weights, the optimal synaptic state is an explicit function of the energy budget (Eqs.~\eqref{eq:closed_forms_ET_rewrite}--\eqref{eq:power_law_rewrite}). \textit{Second}, precision obeys a strict power law $k^{-2}\propto (E^*)^{5}$ (Eq.~\eqref{eq:power_law_rewrite}), so multiplicative budget changes induce predictable gains in reliability. In \hyperref[equivalence_general]{Appendix -- Equivalence of trade–off and energy–budget formulations}, we show that the budget-constrained and tradeoff formulations are equivalent at the optimum, enabling us to treat budgets as the primary control variable for inference and for modelling plasticity driven by budget updates.

\subsection{Plasticity updates the energy budget}
\label{sec:plasticity_updates_budget}

To predict post-plasticity expression we require a rule for how the synaptic energy budget changes from its
baseline value \(E_0\) to \(E_1\). Under the minimal--energy boundary hypothesis, synapses remain close to the
active budget surface \(E=E^*\) throughout plasticity; consequently forecasting
\((n_1,p_1,q_1,\sigma_1^2)\) requires forecasting \(E_1\) itself.

\paragraph{Inferring the baseline budget \(E_0\) and defining a state-based post target \(E_1\).}
We infer the baseline budget from pre--plasticity summary statistics \((\mu_0,\sigma_0^2)\) using the
calcium pump\(+\)turnover closed-form relations (Eqs.~\eqref{eq:closed_forms_ET_rewrite}--\eqref{eq:kappa_def}):
\begin{align}
E_0 \;=\; \kappa(\beta_p)\Big(\tfrac{\mu_0^2}{\sigma_0^2}\Big)^{1/5},
\qquad
\kappa(\beta_p)\;=\;\tfrac{4}{5}\,\beta_p^{-1}\!\big[5(1-\beta_p)\big]^{4/5}.
\label{eq:E0_from_stats}
\end{align}
For held--out evaluation we define a \emph{state-based} post-plasticity budget target
\begin{align}
E_1^{\text{state}}
\;=\;
\beta_p\,b_1^{1/4}+(1-\beta_p)\,n_1,
\qquad
b_1=\frac{p_1}{1-p_1},
\label{eq:E1_state}
\end{align}
which depends only on \((n_1,p_1)\) and \emph{does not use} \(\sigma_1^2\). Throughout we set $\beta_p=0.95$.

\begin{figure}[H]
\centering
\includegraphics[width=1\textwidth]{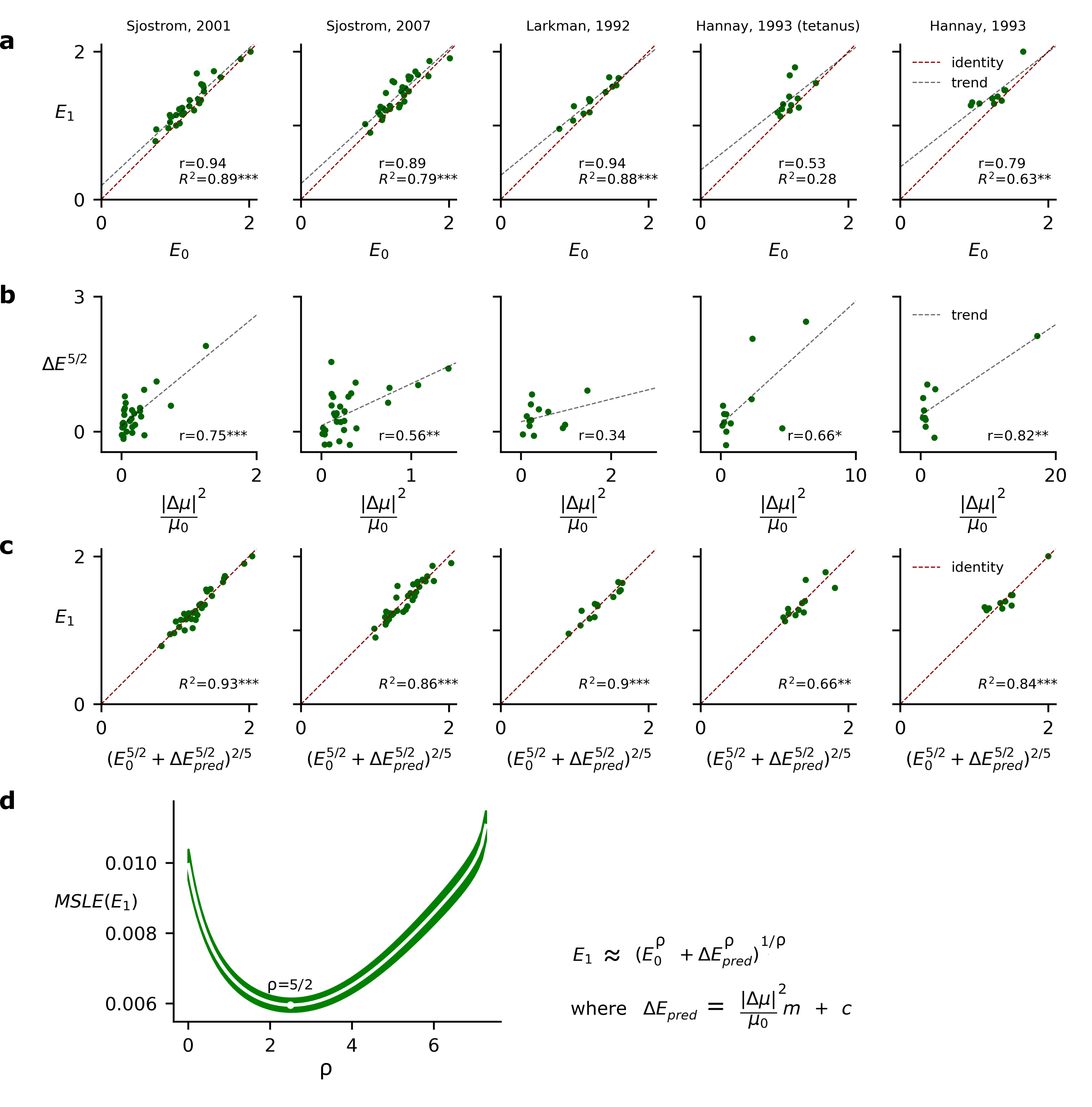}
\caption[Modelling the update of energy budgets.]{
\textbf{Fitting budget updates.}
\textbf{a)} \(E_1\) versus \(E_0\) (dashed: identity). Most points lie above the line, indicating budget
increases after plasticity.
\textbf{b)} Linear relation between \(\Delta E^{5/2}\) and the baseline-adjusted squared mean change
\(|\Delta\mu|^{2}/\mu_0\).
\textbf{c)} Out-of-sample prediction of \(E_1\) using
\(E_1^{\text{pred}}=\big(E_0^{5/2}+\Delta E_{\text{pred}}^{5/2}\big)^{2/5}\)
(dashed: identity).
\textbf{d)} Exponent sweep for the energy coordinate $E^{\rho}$. Mean squared log error (MSLE) of held-out
\(E_{1}^{\text{state}}\) under leave-one-out cross-validation across all datasets (white curve = mean; green
shaded band = $\pm$ SEM). The minimum clusters near $\rho\!\approx\!5/2$. \emph{Note:} the empirically selected
exponent coincides with the pump\(+\)turnover precision--energy law (Eq.~\eqref{eq:power_law_rewrite}), aligning the
update coordinate with the natural reliability coordinate.
}
\label{fig_tuning_sigma}
\end{figure}

\paragraph{Two ingredients: budget persistence and stimulus-dependent reallocation.}
Figure~\ref{fig_tuning_sigma}a shows a strong monotone relationship between \(E_0\) and \(E_1\), indicating that
budgets largely persist across plasticity. This is expected if \(E\) reflects slowly changing structural and
metabolic constraints (e.g.\ presynaptic machinery capacity, calcium-handling infrastructure, local energy
supply), making \(E_0\) a natural baseline predictor of \(E_1\).
However, persistence alone cannot explain the systematic upward shift \(E_1>E_0\) observed for most synapses
(Fig.~\ref{fig_tuning_sigma}a), nor does it specify how much a given synapse's budget should increase.
We therefore seek an additional synapse-local plasticity signal that predicts the increment.

\paragraph{A minimal driver of energy investment.}
A useful driver should be (i) synapse-local; (ii) comparable across synapses with different baseline
strengths; and (iii) predictive of the extra energetic investment beyond persistence.
These requirements motivate the parsimonious quantity
\begin{align}
x \;\equiv\; \frac{|\Delta\mu|^2}{\mu_0},
\qquad \Delta\mu=\mu_1-\mu_0,
\label{eq:driver_x}
\end{align}
which measures the baseline-adjusted squared magnitude of the mean update.
Squaring upweights large events; normalising by \(\mu_0\) compensates for the empirical tendency of larger
baseline synapses to undergo larger absolute changes in \(\Delta\mu\), yielding a driver that is more comparable
across synapses.
(\emph{Note:} since \(\mu\) is measured in mV, \(x\) is not dimensionless; it is best interpreted as a
scale-adjusted local update magnitude.)

\paragraph{Update model family: additivity in an energy coordinate.}
We model the stimulus-driven increment as additive in a transformed energy coordinate,
\begin{align}
E_1^{\rho} \;=\; E_0^{\rho} + m\,x + c,
\qquad
E^{\text{pred}}_1=\big(E_0^{\rho}+m\,x+c\big)^{1/\rho}.
\label{eq:update_family_better}
\end{align}
The rationale is operational but principled: plasticity should enact an approximately \emph{linear} increment
in the coordinate that directly parameterises \emph{precision investment}; since the correct coordinate is not
obvious a priori, we allow the data to identify it.
Concretely, we sweep \(\rho\), refit \(m,c\) for each value,
and select \(\rho\) by minimising out--of--sample mean squared log error (MSLE) for the target
\(E_1^{\text{state}}\) under leave-one-out cross-validation across all datasets (Fig.~\ref{fig_tuning_sigma}d;
see \hyperref[updating_E]{Methods -- Updating the energy budget}).
The optimum clusters near \(\rho\approx 5/2\), so \(\Delta E^{5/2}\) is approximately linear in \(x\)
(Fig.~\ref{fig_tuning_sigma}b), and yields accurate held-out predictions of \(E_1^{\text{state}}\)
(Fig.~\ref{fig_tuning_sigma}c).

\paragraph{Why \(\rho=5/2\) is not just a fit: alignment with the precision--energy law.}
The empirical choice \(\rho\approx 5/2\) is striking because it matches the calcium pump\(+\)turnover theory linking
energy to inverse precision. From Eq.~\eqref{eq:power_law_rewrite},
\begin{align}
k^{-2}=\frac{\mu^{2}}{\sigma^{2}}=\Big(\tfrac{E}{\kappa(\beta_p)}\Big)^{5}
\qquad\Rightarrow\qquad
k^{-1}=\frac{\mu}{\sigma}=\Big(\tfrac{E}{\kappa(\beta_p)}\Big)^{5/2}.
\label{eq:k_E_relation}
\end{align}
Thus, additivity in \(E^{5/2}\) is additivity in the coordinate controlling inverse noise-to-mean.
Substituting \(\rho=5/2\) into Eq.~\eqref{eq:update_family_better} implies an approximately linear precision
shift driven by local plasticity magnitude:
\begin{align}
E_1^{5/2} \;\approx\; E_0^{5/2} + m\,\frac{|\Delta\mu|^2}{\mu_0}+c
\qquad\Rightarrow\qquad
k_1^{-1}\;\approx\; k_0^{-1} + \kappa(\beta_p)^{-5/2}\!\left(m\,\frac{|\Delta\mu|^2}{\mu_0}+c\right),
\label{eq:k_update}
\end{align}
i.e.\ \(\Delta k^{-1}\propto \Delta E^{5/2}\propto |\Delta\mu|^2/\mu_0\) up to constants.
In this sense, the budget update is a synapse-local rule for \emph{precision investment}; synapses that were
changed more by the stimulus receive larger increments in the energy coordinate that directly controls
reliability.

\paragraph{Final budget update rule.}
We therefore adopt the parsimonious update model
\begin{align}
\Delta E^{5/2} \;\approx\; m\,\frac{|\Delta\mu|^{2}}{\mu_0} + c,
\qquad
E_1 \;=\; \big(E_0^{5/2} + \Delta E^{5/2}\big)^{2/5},
\label{eq:final_budget_update}
\end{align}
capturing both slow persistence (via \(E_0\)) and stimulus-dependent increments (via \(|\Delta\mu|^{2}/\mu_0\)).

Having specified how budgets change during plasticity, we now ask whether the inferred post-plasticity
budget \(E_1\) improves prediction of post-plasticity reliability.

\subsection{Energy budgets predict synapse noise}
\label{predicts_noise}

We now test the central empirical prediction of the framework: if plasticity updates an effective synaptic
energy budget, then incorporating the inferred post-plasticity budget \(E_1\) should improve
\emph{out-of-sample} prediction of post-plasticity variance \(\sigma_1^2\).

This test is non-trivial because the budget update rule was fit against a \emph{state-based} target
\(E_1^{\text{state}}(n_1,p_1)\) (Eq.~\eqref{eq:E1_state}) that does \emph{not} use \(\sigma_1^2\).
Accordingly, any improvement in predicting \(\sigma_1^2\) constitutes a genuine consequence of the budget
update rather than circular fitting.

\paragraph{Competing models.}
For each synapse we compare three models for predicting \(\sigma_1^2\) from \((\mu_0,\sigma_0^2,\mu_1)\).
These comprise (i) a fixed-budget persistence control, (ii) the proposed plasticity-driven budget update,
and (iii) a simple variance--mean scaling baseline that preserves \(\sigma^2/\mu\).

\begin{figure}[H]
\centering
\includegraphics[width=1\textwidth]{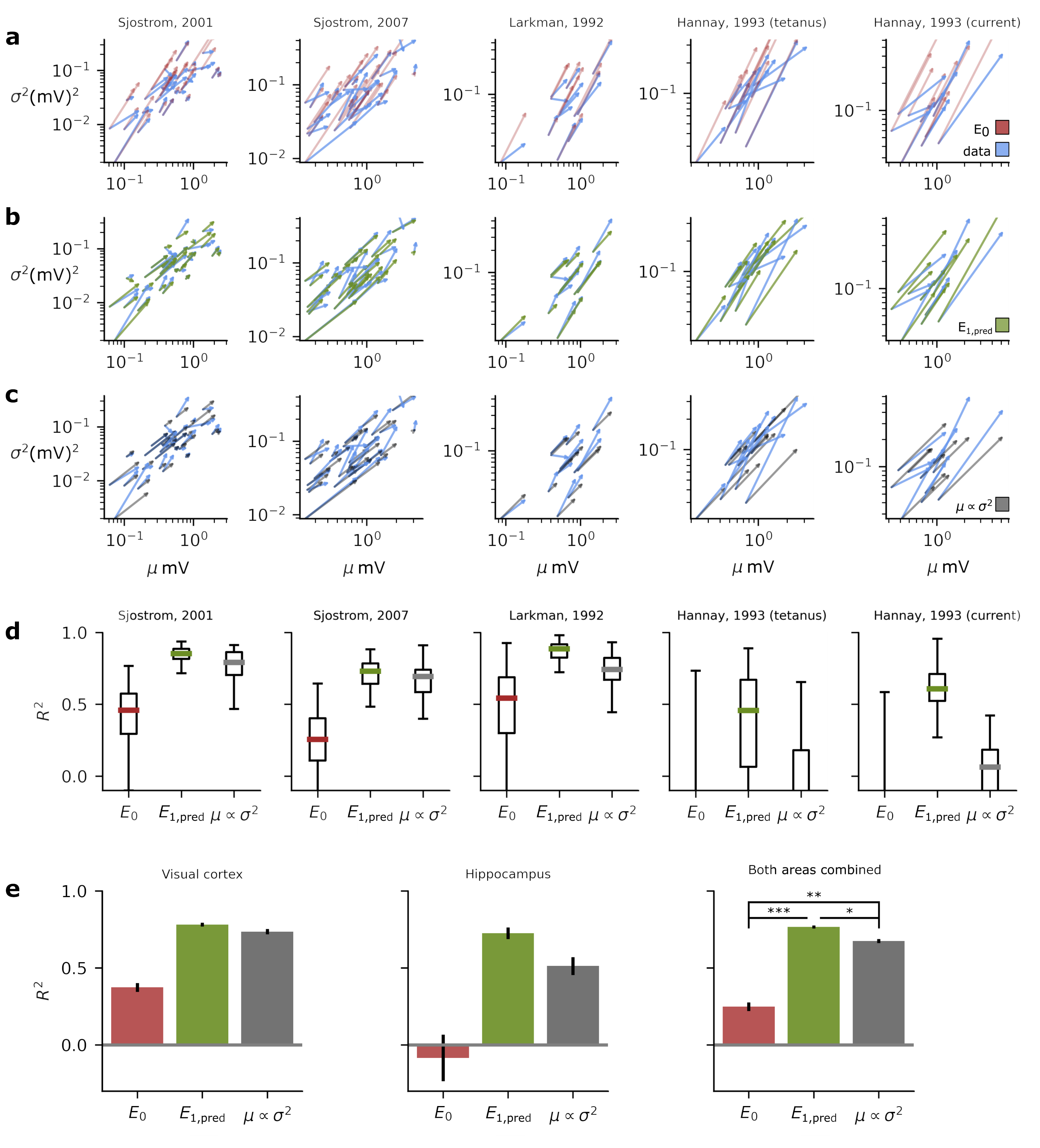}
\caption[Model predictions and validation of synaptic plasticity across datasets.]{
\textbf{Budget updates predict post-plasticity variance.}
\textbf{(a--c)} Predicted changes in $\sigma^2$ under three models. Blue arrows represent observed plasticity updates,
while red, green, and gray arrows represent model predictions. 
\textbf{a)} Fixed budget ($E=E_0$). 
\textbf{b)} Updated budget ($E=E_{1,\text{pred}}$). 
\textbf{c)} Baseline with $\sigma^{2}/\mu$ held constant \citep{song2005highly, aitchison2021synaptic}. 
\textbf{d)} Bootstrap $R^2$ for population-level fits (box: interquartile range; whiskers: 5--95\%). 
\textbf{e)} Cross-validated $R^2$ within each dataset (leave-one-out; bars: mean, error bars: $\pm$\,SEM) for visual cortex \citep{sjostrom2001rate, sjostrom2007multiple}, hippocampus \citep{larkman1992presynaptic, hannay1993common}, and the combined set. 
Asterisks denote one-sided paired tests 
($^{*}$~$p<0.05$, $^{**}$~$p<0.01$, $^{***}$~$p<0.001$). 
The updated-budget model (green) consistently outperforms both alternatives. 
}
\label{fig_updating_sigma}
\end{figure}

\begin{enumerate}[label=(\roman*)]
\item \textbf{Fixed-budget model (\(E=E_0\)).} Budgets do not change across plasticity:
\(E_1^{\text{pred}}=E_0\), where \(E_0\) is inferred from \((\mu_0,\sigma_0^2)\) via Eq.~\eqref{eq:E0_from_stats}.
\item \textbf{Updated-budget model (\(E=E_{1,\text{pred}}\)).} Budgets update according to the fitted rule
(Eq.~\eqref{eq:final_budget_update}), producing \(E_{1,\text{pred}}\) from \(E_0\) and the driver
\(x=|\Delta\mu|^2/\mu_0\) (Eq.~\eqref{eq:driver_x}).
\item \textbf{Constant-\(\sigma^2/\mu\) baseline.} Variance scales proportionally with the mean so that
\(\sigma^2/\mu\) is preserved from pre to post:
\begin{align}
\frac{\sigma_1^{2}}{\mu_1}=\frac{\sigma_0^{2}}{\mu_0}
\qquad\Rightarrow\qquad
\sigma_{1,\text{pred}}^{2}=\sigma_0^{2}\,\frac{\mu_1}{\mu_0}.
\label{eq:constant_sigma2_over_mu}
\end{align}
This baseline is motivated by empirical co-variation between \(\mu\) and \(\sigma^2\) in some regimes
(e.g.\ \citealp{song2005highly, aitchison2021synaptic}), but it omits explicit energy reallocation and, in our
datasets, the proportionality weakens over short plasticity windows (Fig.~\ref{fig_updating_sigma}c--e),
making it a natural but incomplete null model.
\end{enumerate}

\paragraph{From budgets to variance predictions.}
Given a predicted post-plasticity budget \(E_{1,\text{pred}}\) and the observed post-plasticity mean \(\mu_1\),
we predict \(\sigma_1^2\) using the calcium pump\(+\)turnover budget--precision mapping (Eq.~\eqref{eq:power_law_rewrite}),
\begin{align}
\sigma_{1,\text{pred}}^{2}
\;=\;
\mu_1^2\Big(\tfrac{\kappa(\beta_p)}{E_{1,\text{pred}}}\Big)^{5},
\label{eq:sigma1_from_budget}
\end{align}
with \(\beta_p=0.95\) throughout. Thus, the only difference between models (i) and (ii) is whether
\(E_{1,\text{pred}}\) is held fixed at \(E_0\) or updated by Eq.~\eqref{eq:final_budget_update}.

\paragraph{Results.}
Across datasets and brain areas, the updated-budget model provides the most accurate account of variance
updates. Qualitatively, its predicted arrows align more closely with the observed post-plasticity
\((\mu,\sigma^2)\) displacements than either the fixed-budget model or the constant-\(\sigma^2/\mu\) baseline
(Fig.~\ref{fig_updating_sigma}a--c). Quantitatively, both the bootstrap population-level fits
(Fig.~\ref{fig_updating_sigma}d) and leave-one-out cross-validated fits within datasets
(Fig.~\ref{fig_updating_sigma}e) show that updating \(E\) substantially improves predictive power for
\(\sigma_1^2\).

\paragraph{Conclusion.}
These results support the core mechanism: plasticity-induced changes in \(\mu\) are accompanied by systematic
reallocation of energy budgets, and the inferred budget increments are sufficient to explain the observed
reductions (or constrained increases) in post-plasticity variance. If synapses can locally track a proxy for
the scale-free driver \(x=|\Delta\mu|^2/\mu_0\), they can tune their reliability after plasticity in proportion
to the magnitude of the learning update, stabilizing synaptic transmission while allocating metabolic
resources preferentially to synapses most impacted by the plasticity stimulus.

\section{Discussion}

We examined how synapses balance signalling reliability against metabolic expenditure, and how this balance constrains synaptic variability during plasticity. By combining closed-form optimisation with empirical analysis of pre- and post-plasticity datasets, we show that synaptic precision is tightly coupled to an inferred energy budget. This framework links synaptic reliability, metabolic allocation, and plasticity through a quantitative energy--precision relationship, and provides a mechanistic account of how energetic constraints regulate synaptic variability.

\subsection{Reliability--efficiency trade-off in synaptic signalling}

A central result is that synapses operate close to a \emph{minimum-energy boundary} compatible with their mean strength. At fixed mean $\mu$, synaptic variance is constrained by an effective budget $E$, such that increasing precision requires increasing energetic investment. This constraint emerges naturally from an optimisation principle balancing variability against energetic cost, formalised equivalently as a trade-off objective $\T=\sigma^2+\gamma E+\mathrm{const}$ or a budget-constrained Lagrangian (Appendix).

This framework yields monotonic relationships between energy availability and quantal parameters. Increasing budget increases the number of release sites ($n^*\propto E^*$), increases release odds ($\tfrac{p^*}{1-p^*}\propto (E^*)^4$), and reduces variability ($\sigma^{2*}\propto (E^*)^{-5}$). These relationships reflect the underlying geometry of the constrained optimisation problem: higher energetic investment enables more reliable transmission through coordinated adjustments in presynaptic and postsynaptic mechanisms. Importantly, these trends arise directly from energetic constraint rather than requiring independent regulation of quantal parameters. Our results therefore elevate energy budget from a passive correlate of synaptic state to an active constraint governing attainable reliability.

\subsection{Evidence for an effective pump-like and turnover-like cost pairing}

Model comparison identified a two-component energetic structure consisting of a dominant calcium pump--like cost paired with a weaker vesicle-number-dependent turnover cost. This pairing provides a strictly monotonic and separable mapping between energy budget and synaptic precision,
\begin{align}
E^*=\kappa(\beta_p)\Big(\tfrac{\mu^2}{\sigma^2}\Big)^{1/5},
\end{align}
implying that proportional changes in budget produce predictable proportional changes in precision. This separability is critical for reliable budget inference from experimentally measured mean and variance.

Physiologically, the dominant cost component reflects the energetic burden of sustaining high release probability through presynaptic Ca$^{2+}$ influx and ATP-dependent restoration, while the secondary component reflects vesicle maintenance and readiness costs. The inferred weighting $\beta_p\approx 0.95$ indicates that energetic constraints primarily penalise increases in release assurance, while vesicle-based averaging provides a secondary reliability mechanism.

This energetic structure aligns with recent optical quantal analyses showing that release probability accounts for much of the variability in synaptic output, with vesicle number playing a secondary role under physiological conditions \citep{durst2022vesicular}. Within our framework, this asymmetry arises naturally because increasing release probability incurs steep energetic penalties, constraining how reliability improvements are expressed across quantal dimensions.

\subsection{Plasticity as a driver of energy budget updates}

Plasticity events were associated with systematic increases in inferred synaptic energy budgets. Under the precision--energy relationship $\mu^2/\sigma^2 \propto (E^*)^5$, budget increases correspond directly to increased reliability. This suggests a simple organising principle: synapses undergoing larger plasticity-induced mean changes receive greater energetic investment, stabilising future transmission.

This principle closely parallels normative accounts in which synaptic precision reflects synapse importance for learning \citep{mackay1992bayesian,neal2012bayesian,rusakov2020noisy,aitchison2021synaptic,schug2021presynaptic}. In these formulations, parameters that contribute more strongly to learning outcomes are assigned greater precision. Our results provide direct physiological evidence for this principle at the level of individual synapses: plasticity magnitude acts as a local signal governing energetic allocation and hence synaptic reliability.

The scale-free relationship between plasticity magnitude and budget update, $\Delta E^{5/2}\propto |\Delta\mu|^2/\mu_0$, is particularly notable. This quantity directly parameterises inverse noise-to-mean under the inferred energy model, indicating that plasticity acts approximately additively in the coordinate controlling synaptic precision. Larger effective learning events therefore produce proportionally larger increases in energetic investment and reliability.

\subsection{Mechanistic implications of energy-constrained reliability}

Although longitudinal trajectories of quantal parameters were not available for direct validation, the inferred energetic constraints imply specific mechanistic trends. Under increasing energy budget, optimal reliability improvements are achieved through coordinated increases in release probability and vesicle availability, with compensatory adjustments in quantal amplitude to preserve mean strength. The relative contribution of these mechanisms depends on the energetic weighting parameters and the current synaptic state.

These trends arise directly from the structure of the constrained optimisation problem and provide testable predictions for future experiments. In particular, manipulations that alter energetic availability or calcium-handling efficiency should systematically alter attainable synaptic precision and the physiological mechanisms through which reliability improvements are expressed.

\subsection{Energy budgets provide a distinct constraint beyond mean--variance scaling}

Synaptic mean and variance co-vary across preparations, consistent with previous observations \citep{song2005highly,aitchison2021synaptic}. However, budget-based predictions provided substantially improved explanations of post-plasticity variance compared to models based solely on mean scaling. This demonstrates that variance dynamics are not determined by mean changes alone, but are additionally constrained by an underlying energetic resource.

This distinction is particularly evident across different plasticity regimes. Longer-timescale plasticity showed stronger adherence to the predicted energy--precision relationship, consistent with the idea that energetic allocation stabilises transmission over sustained learning periods. Short-timescale plasticity exhibited weaker coupling, suggesting that energetic constraints may be relaxed transiently during rapid synaptic modulation.

\subsection{Limitations and future directions}

Several limitations should be noted. First, energy budgets are inferred indirectly from synaptic statistics and represent effective constraints rather than direct measurements of metabolic expenditure. Second, available datasets provide only pre- and post-plasticity endpoints, limiting direct observation of intermediate physiological adjustments. Third, quantal parameters inferred under binomial assumptions represent effective connection-level descriptions that may compress underlying heterogeneity.

Future experiments combining quantal analysis with targeted metabolic perturbations and time-resolved measurements will be essential for testing the predicted relationships between energy availability and synaptic reliability. Direct manipulation of energetic supply or calcium-handling mechanisms would provide particularly strong tests of the inferred cost structure and precision--energy relationship.

\subsection{Summary}

These results establish synaptic energy budgets as a fundamental constraint governing synaptic reliability and plasticity. By demonstrating that synaptic precision follows a strict and predictable relationship with inferred energetic availability, this work identifies energy allocation as a concrete biophysical mechanism linking metabolic constraint, synaptic variability, and learning. More broadly, these findings provide experimental support for the hypothesis that biological learning systems regulate precision through energy allocation, revealing a quantitative principle underlying synaptic plasticity.

\section{Materials and methods}

\subsection{Synaptic Plasticity Data}
\label{synaptic_plasticity_data}
Datasets were sourced from five published experiments (Table~\ref{table:plasticity}), each reporting
EPSP mean (\(\mu\)) and variance (\(\sigma^2\)) for synapses before and after plasticity induction.
For \citet{sjostrom2001rate} and \citet{sjostrom2007multiple} datasets, the effective binomial quantal parameters \((n,p,q)\) were derived by \citet{costa2017synaptic}, who inferred
them using standard variance--mean / histogram-based quantal analysis under a binomial release model; \(n\) was fixed across synapses
(\(n=5.5\)) using prior estimates of functional release sites for the same connection type
\citep{markram1997physiology}. For the remaining datasets, \citet{larkman1992presynaptic,hannay1993common}, \(n\) was estimated by the original authors from EPSP
histograms. Throughout, these estimates implicitly treat \(n\) as approximately constant over the
recording window; while longer-timescale structural remodelling can alter effective release site number,
available evidence suggests that large changes in \(n\) are limited on sub-hour timescales in these
preparations \citep{bolshakov1997recruitment, saez2009plasticity}.

\begin{table}[H]
\centering
\caption{Summary of plasticity experiments}
\renewcommand{\arraystretch}{1.2}
\setlength{\tabcolsep}{5pt}
\begin{tabular}{@{}p{2.5cm}p{3.5cm}p{5cm}p{4cm}@{}}
\toprule
\textbf{Study} & \textbf{Region/\newline Cell Type} & \textbf{Plasticity description} & \textbf{Measurement} \\ \midrule
Sjostrom, 2001 (n=30) & Visual cortex,\newline L5 neurons &
STDP induced LTP via current \newline injection with varied timing \newline and frequency. Lasted 80 minutes &
Whole-cell patch-clamp \newline recordings \newline Interval: 10--80 minutes \\ \midrule
Sjostrom, 2007 (n=31) & Visual cortex,\newline L5 neurons &
LTP via paired, prolonged \newline high-frequency current injections.\newline Lasted 60 minutes &
Whole-cell recordings \newline Interval: 10--60 minutes \\ \midrule
Larkman, 1992 (n=13) & Hippocampus, CA1 \newline pyramidal neurons &
LTP via presynaptic tetanic \newline stimulation and additional \newline fiber tetanus. Lasted $>$ 15 minutes &
Sharp electrode \newline recordings \newline Interval: 15 minutes \\ \midrule
Hannay, 1993 (Tetanus) (n=11) & Hippocampus, CA1 \newline pyramidal neurons &
STP via tetanic stimulation; \newline EPSPs decayed within 6 minutes; \newline possible contribution of PTP due to residual calcium buildup &
Sharp electrode \newline recordings \newline Interval: 1 minute \\ \midrule
Hannay, 1993 (Current) (n=10) & Hippocampus, CA1 \newline pyramidal neurons &
STP via presynaptic stimulation \newline and 40-second postsynaptic \newline current injection. Decayed within 6 minutes &
Sharp electrode \newline recordings \newline Interval: 1 minute \\ \bottomrule
\end{tabular}
\label{table:plasticity}
\end{table}

A potential concern is whether these datasets reflect compound responses from multiple connections.
For \citet{sjostrom2001rate,sjostrom2007multiple}, EPSPs were measured using paired whole-cell recordings
between identified pyramidal-cell pairs, and therefore reflect transmission for a single presynaptic neuron
onto a single postsynaptic neuron (a unitary connection). Such connections typically comprise multiple
presynaptic boutons contacting the postsynaptic cell \citep{markram1997physiology,silver2003estimation}.
Accordingly, \(n\) indexes the number of functional release sites within the unitary connection; \(p\)
denotes a connection-average release probability (acknowledging bouton-level heterogeneity); and \(q\) is
an effective quantal size.

By contrast, the hippocampal datasets of \citet{larkman1992presynaptic} and \citet{hannay1993common} used
minimal extracellular stimulation, a standard approach to activate a single (or very few) afferent fibres
and obtain small, fluctuating EPSPs suitable for quantal analysis \citep{faber1991applicability}. Although
minimal stimulation cannot absolutely guarantee single-fibre isolation, multiple studies show that the
resulting amplitude distributions, failure rates, and variance--mean structure are consistent with
activation of single or very few axons in CA1 \citep{hessler1993probability,rosenmund1996definition}.
Accordingly, these preparations have long been validated and used for quantal analysis in hippocampal CA1
\citep{herreras1987characteristics,enoki2009expression}. Across both preparations, binomial variance--mean
analysis is standard practice for estimating effective quantal parameters despite underlying bouton
heterogeneity \citep{faber1991applicability,enoki2009expression,costa2017synaptic}.

Finally, while release probability (\(p\)) can vary across individual boutons within the same connection
\citep{murthy1997heterogeneous,enoki2009expression}, theoretical and empirical work suggests that treating
\(p\) as uniform provides a good first-order approximation for connection-level mean and variance
\citep{faber1991applicability,silver2003estimation,loebel2009multiquantal}. We therefore interpret the
inferred binomial parameters \(n,p,q\) throughout as \emph{effective connection-level values}: that is, coarse-grained parameters that summarise the collective
release statistics of multiple heterogeneous boutons within a unitary connection.
Under this interpretation, \(n\) reflects the effective number of functional release sites, \(p\) the
connection-averaged release probability, and \(q\) an effective quantal amplitude that incorporates both
vesicle content and postsynaptic responsiveness.
These parameters are not assumed to be identical across boutons, but rather to capture their aggregate
impact on the observed mean and variance of transmission.

\subsection{Energy cost selection and rationale}
\label{methods_energy_cost}

Our framework requires an explicit mapping from quantal state $\theta=(n,p,q)$ to metabolic expenditure.
Rather than assuming a unique ground-truth energy function, we specify a small set of candidate
physiological cost components motivated by scaling arguments identified in \citet{malkin2024signatures}, and then select the best-supported effective
energy model empirically. Selection is based on which parameterisations most consistently reproduce
observed quantal states under our minimum-energy boundary hypothesis.

\subsubsection{Candidate cost components and dependence on (n,p,q)}
\label{subsec:cost_components}

We define five candidate cost components $C_i(\theta)$, each intended to penalise a distinct reliability
strategy---the first four taken from \citet{malkin2024signatures}. These are not meant as a complete molecular accounting, but as a parsimonious basis set whose
relative weights can be inferred.

\paragraph{(1) Presynaptic calcium pumping cost, $C_p(p)$.}
Release probability depends steeply on presynaptic Ca$^{2+}$, often approximated by a Hill relationship
with exponent $\eta\approx 4$ \citep{sakaba2001quantitative}. Writing release odds $b=p/(1-p)$,
$b\propto [{\rm Ca}]^\eta$ implies $[{\rm Ca}]\propto b^{1/\eta}$. Assuming energetic expenditure scales
with Ca$^{2+}$ extrusion/sequestration per event yields
\begin{equation}
C_p(p) \propto b^{1/\eta} \approx \left(\frac{p}{1-p}\right)^{1/4}.
\label{eq:cp}
\end{equation}
This term rises sharply as $p\to 1$, capturing the intuition that high release probability is expensive.

\paragraph{(2) Vesicle membrane maintenance / proton leak cost, $C_m(n,q)$.}
Increasing $n$ reduces release noise by averaging but increases vesicle membrane area and associated
maintenance costs (lipid turnover, reversal of proton leak) \citep{purdon2002energy,pulido2021synaptic}.
If quantal size scales with vesicle volume (see \citet{karunanithi2002quantal}), $q\propto r^3$, then
surface area scales as $r^2\propto q^{2/3}$, giving
\begin{equation}
C_m(n,q) \propto n\,q^{2/3}.
\label{eq:cm}
\end{equation}

\paragraph{(3) Actin/cytoskeletal scaffolding cost, $C_a(n,q)$.}
A simple scaling assumes actin support per vesicle scales with diameter ($\propto r$), giving
\begin{equation}
C_a(n,q) \propto n\,q^{1/3}.
\label{eq:ca}
\end{equation}

\paragraph{(4) Vesicle trafficking / mobilisation cost, $C_{tr}(n,p)$.}
A minimal activity-dependent mobilisation cost scales with the expected released vesicle count per spike,
\begin{equation}
C_{tr}(n,p) \propto n p.
\label{eq:ctr}
\end{equation}

\paragraph{(5) Protein turnover / readiness cost, $C_t(n)$.}
Maintenance and replacement of vesicle-associated and release machinery scales with the vesicle pool size,
motivating
\begin{equation}
C_t(n) \propto n.
\label{eq:ct}
\end{equation}

\subsubsection{Energy model class}
\label{subsec:energy_model_class}

Total energy is modelled as a convex combination of candidate components,
\begin{align}
E(\theta;\bm\beta)=\sum_i \beta_i\,C_i(\theta),
\qquad
\beta_i\ge 0,\quad \sum_i \beta_i=1,
\qquad
\theta=(n,p,q),
\label{eq:E_beta_methods}
\end{align}
where $\beta_i$ are effective weights encoding the relative contribution of each physiological cost under
the reduced model.

\subsubsection{Two-stage constrained optimisation}
\label{subsec:two_stage_fit}

For each synapse with observed \emph{baseline} summary statistics $(\mu_d,\sigma_d^2)$ and for each
candidate $\bm\beta$, we operationalise the minimum-energy boundary hypothesis using a two-stage
procedure.
Throughout this section, $(\mu_d,\sigma_d^2)$ refer to \emph{pre-plasticity} measurements; post-plasticity
quantal parameters are not used in model fitting or selection.

\begin{enumerate}[label=(\roman*)]
\item \textbf{Minimal-energy boundary estimation (budget inference).}
We compute the minimal feasible energy compatible with the observed mean and variance,
\begin{align}
E_d^{*}(\bm\beta)
=
\min_{\theta}\ E(\theta;\bm\beta)
\quad \text{s.t.}\quad
\mu(\theta)=\mu_d,\ \sigma^{2}(\theta)=\sigma_d^{2}.
\label{eq:stage1_methods}
\end{align}

\item \textbf{Noise-minimising state prediction on the boundary.}
With the budget fixed to $E=E_d^{*}(\bm\beta)$, we predict the energy-efficient state by minimising
variance at fixed mean,
\begin{align}
\theta_d^{*}(\bm\beta)
=
\arg\min_{\theta}\ \sigma^{2}(\theta)
\quad \text{s.t.}\quad
\mu(\theta)=\mu_d,\ E(\theta;\bm\beta)=E_d^{*}(\bm\beta),
\label{eq:stage2_methods}
\end{align}
yielding predicted $(n^*,p^*,q^*,\sigma^{2*})$ for each synapse.
\end{enumerate}

Intuitively, Stage~(i) assigns each synapse the smallest budget consistent with its observed baseline
reliability; Stage~(ii) predicts how that budget should be allocated across quantal degrees of freedom
if the synapse selects states that minimise variability under energetic constraint.
Importantly, this procedure uses only baseline statistics and does not condition on post-plasticity
quantal expression.

\subsubsection{Search strategy over model parameterisations}
\label{subsec:beta_search_strategy}

We evaluated two complementary families of parameterisations:

\begin{enumerate}[label=(\alph*)]
\item \textbf{Full-model weight search (coarse simplex grid).}
To assess robustness across all components simultaneously, we performed an exhaustive grid search over the
simplex $\sum_i\beta_i=1$ with step size $\Delta\beta=0.1$, yielding $10^5$ candidate $\bm\beta$ settings
(Fig.~\ref{fig_parallel}).

\item \textbf{Two-cost pairing sweeps (fine edge search).}
To identify which $n$-dependent cost best complements the calcium pump term, we restricted to pairings of
the form $E=\beta_p C_p + (1-\beta_p)C_j$ and swept $\beta_p\in[0,1]$ in steps of $0.01$
(Fig.~\ref{fig_nrmse}).
\end{enumerate}

\subsubsection{Error metric}
\label{subsec:energy_cost_metric}

Predictive accuracy was quantified using the normalised root mean squared error (NRMSE),
\begin{align}
\mathrm{NRMSE}(\theta^*)
=
\frac{\sqrt{\frac{1}{M}\sum_{m=1}^{M}(\theta^{*}_{m}-\theta_{m})^{2}}}{\mathrm{mean}(\theta)},
\qquad
\theta\in\{n,p,q,\sigma^{2}\},
\label{eq:nrmse_methods}
\end{align}
where $\theta$ denotes empirical measurements across $M$ synapses and $\theta^*$ the corresponding
predictions. We report NRMSE for $\sigma^2$ and for joint $(n,p,q)$ prediction.

\subsubsection{Numerical optimisation}
\label{subsec:energy_cost_optimisation}

Optimisation was performed using the \texttt{trust-constr} algorithm (\texttt{scipy.optimize.minimize},
Python 3.10). Objectives were optimised over $(n,p,q)$ subject to the equality constraints above with bounds
\begin{align}
n \ge 1,\qquad 0 \le p \le 1,\qquad q \ge 0,
\end{align}
and initialised using empirical estimates where available. Convergence was declared when the relative
gradient norm fell below $10^{-6}$.

\subsubsection{Results of energy model selection}
\label{subsec:energy_cost_selection_outcome}

\paragraph{Full-model grid search.}

\begin{figure}[H]
    \centering
    \includegraphics[width=0.78\textwidth]{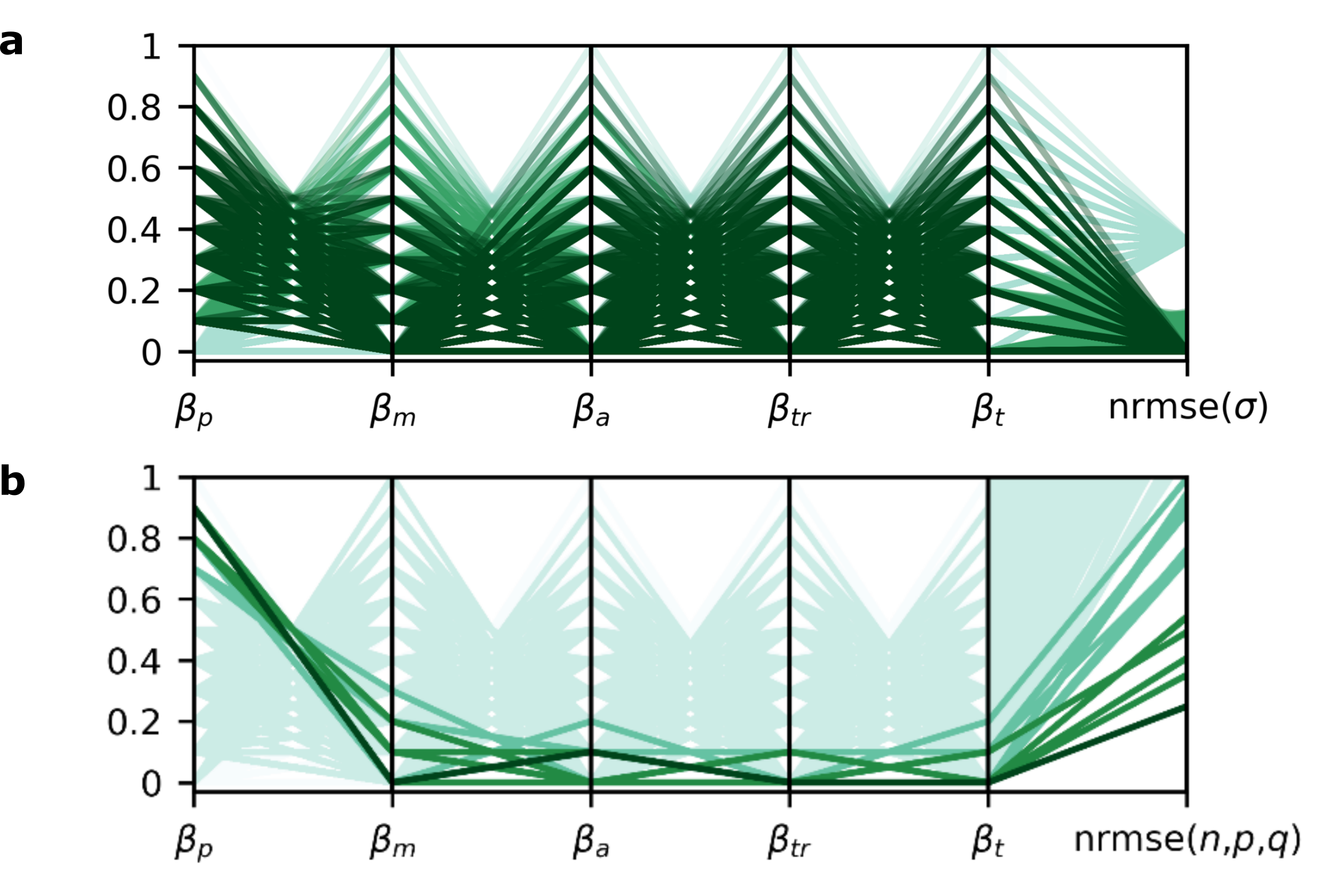}
    \caption[Energy-cost model selection across the full simplex]{\textbf{Energy-cost model selection across the full simplex.}
    Parallel coordinate plots summarise NRMSE between predicted and observed synaptic states across an
    exhaustive grid of cost weights $\bm\beta$ (step size $0.1$, $10^5$ settings). Each polyline corresponds
    to one setting of $\bm\beta$ (intersections at $\beta_i$) and terminates at its NRMSE score. Darker lines
    indicate lower NRMSE (better fit). \textbf{a)} Variance predictions are robust across many
    parameterisations provided $\beta_p>0$ and $\beta_p\neq 1$. \textbf{b)} Accurate prediction of $(n_0,p_0,q_0)$
    requires a dominant calcium pump contribution (typically $\beta_p\gtrsim 0.7$).}
    \label{fig_parallel}
\end{figure}
Figure~\ref{fig_parallel} summarises performance over the full simplex grid.
Two qualitative conclusions are robust.

\emph{First}, accurate modelling of variance requires a non-zero calcium pump contribution ($\beta_p>0$)
and at least one additional $n$-dependent cost ($\beta_p\neq 1$). If the only cost is $C_p(p)$, then under
variance minimisation at fixed $\mu$ the model drives $n\to\infty$ (and $q\to 0$) to suppress
$\sigma^2/\mu^2=\tfrac{1-p}{np}$ without bound. Any non-zero cost that increases with $n$ prevents this
degeneracy and yields stable predictions.

\emph{Second}, while $\sigma^2$ predictions are relatively robust once these conditions are met, accurate
prediction of quantal expression $(n,p,q)$ is more selective: high-performing parameterisations consistently
place a large share on calcium pumping (typically $\beta_p\gtrsim 0.7$), with the remaining weight
distributed across $n$-related costs. This indicates an effective regime in which calcium handling dominates
the energetic constraint shaping synaptic precision, while vesicle-associated costs act as necessary
secondary regularisers.

\begin{figure}[H]
    \centering
    \includegraphics[width=1\textwidth]{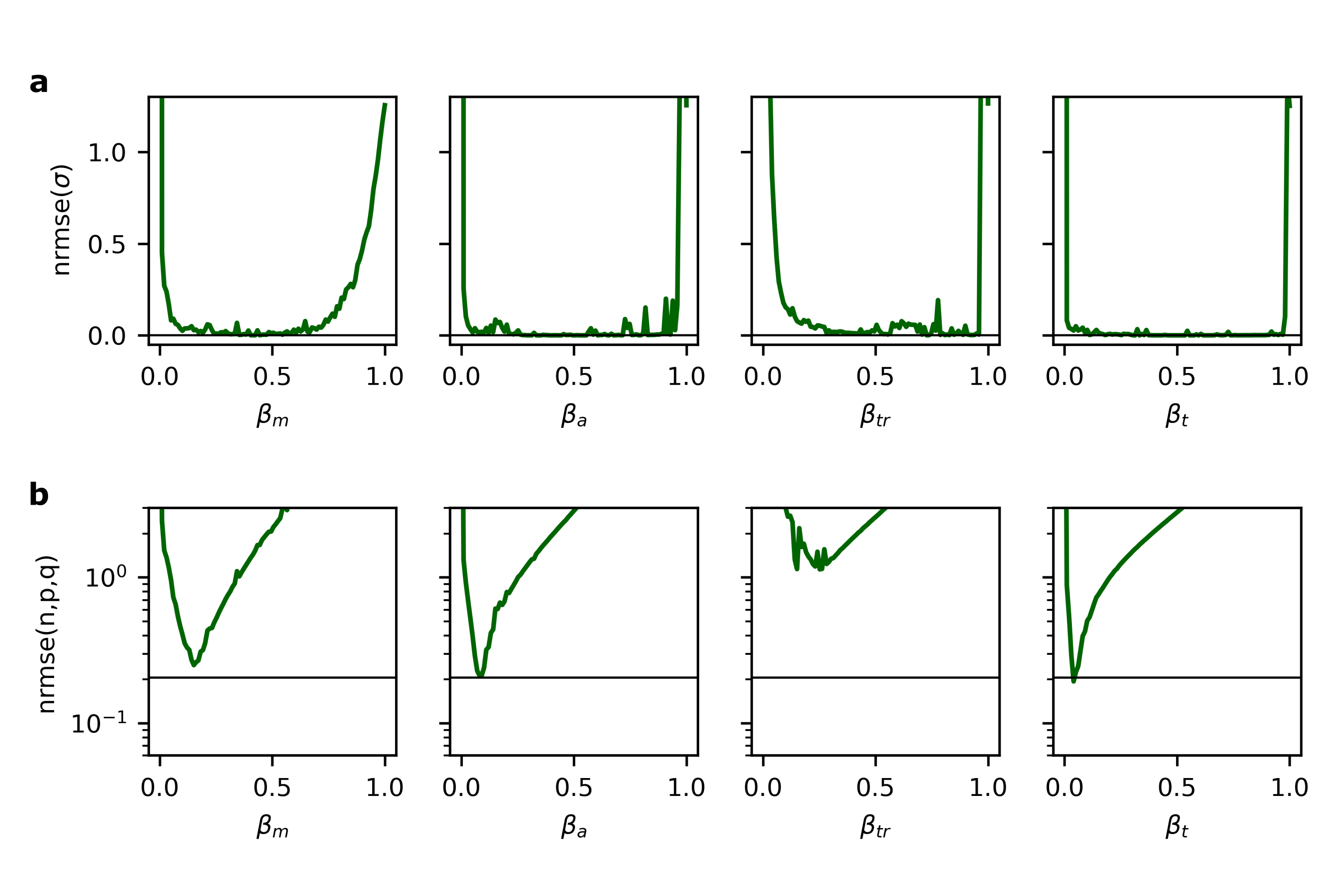}
    \caption[Pairwise comparison of calcium pump plus one additional cost]{\textbf{Pairwise comparison of calcium pump plus one additional cost.}
    NRMSE for predictions of $\sigma_0^{2}$ (\textbf{a}) and $(n,p,q)$ (\textbf{b}) for pairwise models combining
    the calcium pump cost with a single $n$-related cost (Eq.~\eqref{eq:pairwise_methods}). The free weight
    $\beta_p$ is swept in steps of $0.01$ with the complementary weight $(1-\beta_p)$ placed on the second
    component. The black horizontal line indicates the best NRMSE achieved in the full simplex grid search
    (Fig.~\ref{fig_parallel}).}
    \label{fig_nrmse}
\end{figure}

\paragraph{Pairwise cost comparisons.}
To isolate which $n$-dependent cost best complements the calcium pump term, we compared two-cost models of
the form
\begin{align}
E(\theta;\beta_p)=\beta_p\,C_p(\theta)+(1-\beta_p)\,C_j(\theta),
\label{eq:pairwise_methods}
\end{align}
where $C_j$ is one of the candidate $n$-related costs. Figure~\ref{fig_nrmse} shows NRMSE as $\beta_p$ is
swept (step size $0.01$).

Across pairings, several $n$-costs yield comparable NRMSE for $\sigma^2$ and reasonable $(n,p,q)$ fits.
However, the calcium pump$+$turnover pairing is distinguished by two practical advantages: (i) it produces
consistently strong joint fits and (ii) it yields a strictly monotone, separable mapping between energy and
precision at fixed $\mu$ (Eq.~\eqref{eq:power_law_rewrite}), which makes subsequent budget inference and
plasticity-driven budget updates analytically transparent.

\paragraph{Selected cost model.}
Taken together, this analysis supports three constraints on plausible effective energy models:
(i) the calcium pump cost must be non-zero; (ii) at least one $n$-dependent cost must contribute to prevent
the $n\to\infty$ degeneracy; and (iii) the best fits to quantal expression favour a pump-dominated regime.
We therefore adopt calcium pump$+$turnover as the energy model for subsequent analyses. Unless
otherwise stated, we use the effective fitted parameterisation
\begin{align}
E = 0.95\,C_p + 0.05\,C_t,
\label{eq:baseline_energy}
\end{align}
noting that $\bm\beta$ should be interpreted as effective weights in a reduced biophysical model rather
than precise molecular energy fractions.

\subsection{Updating the energy budget}
\label{updating_E}

To predict post-plasticity synaptic states under the minimum--energy boundary hypothesis, we require a rule
for how the active energy budget shifts across plasticity from $E_0$ to $E_1$.
We treat the budget as a latent control variable, not directly measured, but inferred from
quantal statistics using the selected energy model (calcium pump+turnover).

\paragraph{Inferring the baseline budget $E_0$.}
For each synapse we infer its pre-plasticity budget from $(\mu_0,\sigma_0^2)$ using the pump+turnover
closed-form mapping on the fixed--mean surface (Results, Eq.~\eqref{eq:power_law_rewrite}):
\begin{align}
E_0 \;=\; \kappa(\beta_p)\Big(\tfrac{\mu_0^2}{\sigma_0^2}\Big)^{1/5},
\qquad
\kappa(\beta_p)\;=\;\tfrac{4}{5}\,\beta_p^{-1}\!\big[5(1-\beta_p)\big]^{4/5}.
\end{align}
This is the minimal feasible energy consistent with the observed baseline summary statistics under the
selected cost pairing.

\paragraph{State-based post-plasticity target $E_1^{\text{state}}$.}
To train and evaluate an update rule we require a post-plasticity target budget that does not depend on
$\sigma_1^2$, so that variance prediction remains a downstream out-of-sample test.
We therefore define a \emph{state-based} target using only the post-plasticity presynaptic state estimates
$(n_1,p_1)$:
\begin{align}
E^{\text{state}}_1 \;=\; \beta_p\,b_1^{1/4} + (1-\beta_p)\,n_1,
\qquad
b_1=\frac{p_1}{1-p_1}.
\end{align}
This target corresponds to the pump+turnover energy of the empirically estimated post state and depends
only on $(n_1,p_1)$ (and $\beta_p$), not on $\sigma_1^2$.

\paragraph{Update model family: additive increments in an energy coordinate.}
We model budget updates using a persistence component and a stimulus-dependent increment. Let
\begin{align}
x \;\equiv\; \frac{|\Delta\mu|^2}{\mu_0}
\end{align}
denote the scale-free magnitude of the mean update. In an energy coordinate $E^\rho$ we assume an
approximately linear increment:
\begin{align}
\Delta E^\rho \;\equiv\; E_1^\rho - E_0^\rho \;\approx\; m\,x + c,
\qquad
E_1^{\text{pred}} \;=\; \big(E_0^\rho + \Delta E^\rho\big)^{1/\rho},
\label{eq:E_update_rule}
\end{align}
where $(m,c)$ are fitted by ordinary least squares on training data.

\paragraph{Exponent selection.}
We do not fix $\rho$ a priori. Instead, we sweep $\rho$ over a grid and refit $(m,c)$ at each value.
We select $\rho$ by minimising out-of-sample prediction error for $E_1^{\text{state}}$ under cross-validation
(Fig.~\ref{fig_tuning_sigma}d). This tests whether the data support the theoretically suggested precision
coordinate ($\rho=5/2$, from $k^{-1}\propto E^{5/2}$) without presupposing it.

\paragraph{Cross-validation protocol and metrics.}
We use leave-one-out cross-validation (LOOCV) over synapses. For each held-out synapse we:
(i) compute $E_0$ from $(\mu_0,\sigma_0^2)$;
(ii) fit $(m,c)$ on the remaining synapses for a fixed $\rho$ using Eq.~\eqref{eq:E_update_rule};
(iii) predict $E_1^{\text{pred}}$ for the held-out synapse;
(iv) compare $E_1^{\text{pred}}$ to its $E_1^{\text{state}}$ target.
We report coefficient of determination ($R^2$) and mean squared log error (MSLE) between predicted and target
budgets. MSLE is used as the selection criterion for $\rho$ in Fig.~\ref{fig_tuning_sigma}d, and we report
mean $\pm$ SEM across held-out folds.

\subsection{Predicting post--plasticity variability and model comparison}
\label{method_variance_preds}

\paragraph{Predicting post--plasticity variance from predicted budgets.}
Given a predicted post-plasticity budget $E^{\text{pred}}_{1}$ from the update rule
(\hyperref[sec:plasticity_updates_budget]{Results -- Plasticity updates the energy budget}), we convert budget to reliability using the pump+turnover
precision--energy law (Eq.~\eqref{eq:power_law_rewrite}),
\begin{align}
k^{-1}_1 \;=\; \kappa(\beta_p)\,\big(E^{\text{pred}}_{1}\big)^{5/2},
\label{eq:k_pred_from_E}
\end{align}
and then predict post-plasticity variance using the identity $k^{-1}=\mu/\sigma$:
\begin{align}
\sigma_{1}^{2\,\text{pred}}
\;=\;
\frac{\mu_{1}^{2}}{\big(k^{-1}_{1}\big)^{2}}
\;=\;
\frac{\mu_{1}^{2}}{\kappa(\beta_p)^{2}\,\big(E^{\text{pred}}_{1}\big)^{5}}.
\label{eq:sigma_pred_from_E}
\end{align}
Crucially, $\sigma_1^2$ is not used in any step of the budget prediction or variance prediction
pipeline; it is reserved for evaluation.

\paragraph{Baseline comparison models.}
We compare the updated-budget model to two natural null models that do not implement
stimulus-dependent reallocation:
\begin{itemize}
\item \textbf{Fixed-budget control:} $E_1=E_0$ (no post-plasticity budget update).
Using Eq.~\eqref{eq:sigma_pred_from_E} with $E^{\text{pred}}_1=E_0$ yields
\begin{align}
\sigma_{1}^{2\,\text{pred}}
\;=\;
\frac{\mu_{1}^{2}}{\kappa(\beta_p)^{2}\,E_{0}^{5}}.
\end{align}

\item \textbf{Constant $\sigma^2/\mu$ control:} preserves the pre--post proportionality surrogate
\begin{align}
\frac{\sigma_1^{2}}{\mu_1}=\frac{\sigma_0^{2}}{\mu_0}
\;\;\Rightarrow\;\;
\sigma_1^{2\,\text{pred}}=\mu_1\,\frac{\sigma_0^{2}}{\mu_0},
\end{align}
motivated by reported mean--variance covariation in synaptic weights
\citep{song2005highly,aitchison2021synaptic}.
\end{itemize}
Together these baselines test whether variance updates can be explained by (i) persistence of budgets
or (ii) mean-driven scaling alone.

\paragraph{Cross-validation protocol.}
Comparisons (Fig.~\ref{fig_updating_sigma}e) are performed using leave-one-out cross-validation (LOOCV) over synapses within each
dataset. On each fold we:
(i) fit the energy-increment regression on the $N-1$ training synapses,
\begin{align}
\Delta E^{5/2} \;\approx\; m\,\frac{|\Delta\mu|^{2}}{\mu_0}+c,
\end{align}
(ii) predict the held-out budget using
\begin{align}
E^{\text{pred}}_{1}=\big(E_{0}^{5/2}+\Delta E_{\text{pred}}^{5/2}\big)^{2/5},
\end{align}
and (iii) generate the held-out variance prediction using Eq.~\eqref{eq:sigma_pred_from_E}.
The same held-out folds are used for all baseline models to ensure paired comparisons.

\paragraph{Metrics and hypothesis tests.}
For each model we compute prediction accuracy using the coefficient of determination ($R^2$).
To compare models at the synapse level, we compute per-synapse squared errors
$e_i=(\sigma_{1,i}^{2\,\text{pred}}-\sigma_{1,i}^2)^2$, form paired differences
$\Delta_i=e_i^{A}-e_i^{B}$, and test $\mathbb{E}[\Delta_i]=0$ using a paired $t$-test (Fig.~\ref{fig_updating_sigma}e (right)).

\paragraph{Bootstrap population summaries.}
For population-level summaries (Fig.~\ref{fig_updating_sigma}d), we additionally estimate uncertainty in $R^2$
via bootstrap resampling over synapses (10{,}000 replicates). For each replicate we resample synapses
with replacement, refit the update regression within the replicate, and evaluate predictions on the
out-of-bag subset to avoid optimistic bias. The resulting bootstrap distribution is reported as median,
interquartile range, and 5--95\% whiskers.

\section{Acknowledgments}
\bibliography{biblio}
\appendix

\begin{appendixbox}
\subsection{The objectives}
\label{objectives}
Biological systems face a fundamental challenge in optimising synaptic plasticity to achieve both accurate performance and energy efficiency. Synaptic weights must be adjusted in a way that balances the trade-off between minimising error in neural computations and conserving the energetic resources required to maintain synaptic structure and function. 

To formalise this balance, we consider two complementary formulations:
\begin{itemize}
    \item \textbf{Trade-off Formulation:} This approach directly incorporates the trade-off between performance and energy costs into a single objective function, \(\T\), and minimises it to find optimal synaptic parameters (\hyperref[tradeoff_section]{Results -- Reliability--efficiency tradeoff}).
.
    \item \textbf{Energy Budget Formulation:} This formulation introduces explicit constraints on the total energy budget \(E^*\), embedding these constraints into the optimisation process via a Lagrangian function, \(\L\) (\hyperref[budget_section]{Results -- Energy budgets}).
\end{itemize}

The trade-off formulation provides a direct perspective on the interplay between performance and energy costs, reflecting the intrinsic biological tradeoffs. In contrast, the energy budget formulation explicitly incorporates resource constraints, offering a framework to model scenarios where energy availability is limited or externally regulated.

In the following sections, we detail these formulations, their underlying assumptions, and the mathematical frameworks used to derive biologically relevant solutions.
\subsubsection{Trade-off formulation}
\label{Tradeoff_formulation}
We assume that the brain learns by adjusting synaptic weights such that the updated mean \(\mu^*\) and variance \(\sigma^{2*}\) optimise a biological goal while minimising energetic costs. This process implies the minimisation of the objective \(\T\), defined as:
\begin{align}
\T = \langle \text{Performance cost} \rangle + \gamma \cdot \text{Energy cost}.
\end{align}
The performance cost term is given by the squared error between the weight \(w\) and the target mean \(\hat{\mu}\):
\begin{align}
\langle \text{Performance cost} \rangle = \langle (w - \hat{\mu})^2 \rangle,
\end{align}
where \(\langle \cdot \rangle\) denotes the expectation \(E[\cdot]\). Expanding this term:
\begin{align}
\langle (w - \hat{\mu})^2 \rangle &= \langle w^2 - 2w\hat{\mu} + \hat{\mu}^2 \rangle, \\
&= \langle w^2 \rangle - 2\langle w \rangle \hat{\mu} + \hat{\mu}^2, \\
&= \sigma^2 + \mu^2 - 2\mu \hat{\mu} + \hat{\mu}^2,
\end{align}
where \(\sigma^2 = \langle w^2 \rangle - \mu^2\) is the variance of \(w\), and \(\mu = \langle w \rangle\) is its mean.

Adding the energy cost term \(\gamma E\), we obtain the total cost:
\begin{align}
\T &= \sigma^2 + (\mu - \hat{\mu})^2 + \gamma E.
\end{align}

Here, \(\hat{\mu}\) represents the mean that minimises the performance cost alone, while \(\mu^*\) represents the mean that minimises the total cost \(\T\). In general, \(\mu^* \neq \hat{\mu}\), as \(\mu^*\) reflects the trade-off between performance and energy efficiency. When modelling plasticity, we assume that the observed \(\mu\) corresponds to \(\mu^*\), the value that minimises \(\T\). 

Substituting \(\mu = \mu^*\), the objective simplifies to:
\begin{align}
\T = \sigma^2 + \gamma E + \text{const},
\end{align}
where \(\text{const} = (\mu^* - \hat{\mu})^2\) is a fixed offset determined by the difference between \(\mu^*\) and \(\hat{\mu}\).

\subsubsection{Budget formulation: Lagrangian}
\label{budget_formulation}
The budget view can be expressed as a constrained optimisation using the Lagrangian
\begin{align}
\mathcal L
\;=\;
\sigma^2
\;+\;
\lambda\,(E-E^*)
\;+\;
\alpha\,(\mu-\mu^*),
\label{eq:L_def}
\end{align}
with multipliers $\lambda$ and $\alpha$. 

The budget problem (minimise $\sigma^2$ at fixed $\mu^*$ and fixed $E^*$) and the tradeoff problem (minimise $\sigma^2+\gamma E$ at fixed $\mu^*$) share the same stationary points: at the shared optimum the multipliers coincide, $\lambda^*=\gamma$. Intuitively, $\gamma$ is the \emph{shadow price} of energy: it equals the marginal increase in objective per marginal relaxation of the energy constraint. This equivalence (see \hyperref[equivalence_general]{Equivalence of trade–off and energy–budget formulations}) justifies switching freely between the two parameterisations depending on whether one wishes to treat energy as a penalty or as a budget.

\subsubsection{Equivalence of trade–off and energy–budget formulations}
\label{equivalence_general}

Consider the trade–off objective
\begin{align}
\T(\theta)\;=\;\sigma^2(\theta)\;+\;\gamma\,E(\theta)\;+\;\text{const},
\qquad \text{with } \mu(\theta)=\mu^{*},
\end{align}
and the budgeted Lagrangian
\begin{align}
\L(\theta)\;=\;\sigma^2(\theta)\;+\;\lambda\big(E(\theta)-E^{*}\big)\;+\;\alpha\big(\mu(\theta)-\mu^{*}\big),
\end{align}
with the \emph{same} $\mu$–constraint and any energy combination that includes the calcium–pump term.

\paragraph{Stationarity.}
For $\theta\in\{n,p,q\}$,
\begin{align}
\partial_{\theta}\T \;=\; \partial_{\theta}\sigma^2\;+\;\gamma\,\partial_{\theta}E\;+\;\alpha^{*}\partial_{\theta}\mu \;=\;0,
\qquad
\partial_{\theta}\L \;=\; \partial_{\theta}\sigma^2\;+\;\lambda^{*}\partial_{\theta}E\;+\;\alpha^{*}\partial_{\theta}\mu \;=\;0.
\end{align}
Thus the stationarity conditions coincide at the same $(n^{*},p^{*},q^{*})$ iff $\lambda^{*}=\gamma$.

\paragraph{KKT for the budget problem.}
We treat the budget as an \emph{inequality} $E(\theta)\le E^{*}$ with multiplier $\lambda\ge 0$. The KKT conditions are
\begin{align}
\lambda^{*}\!\ge 0,\quad E(\theta^{*})\le E^{*},\quad \lambda^{*}\big(E(\theta^{*})-E^{*}\big)=0,\quad
\partial_{\theta}\sigma^2+\lambda^{*}\partial_{\theta}E+\alpha^{*}\partial_{\theta}\mu=0.
\end{align}
Hence:
\begin{align}
\lambda^{*}\!>\!0 \Rightarrow E(\theta^{*})=E^{*}\ \text{(active)}, 
\qquad
E(\theta^{*})<E^{*} \Rightarrow \lambda^{*}\!=\!0\ \text{(slack)}.
\end{align}

\paragraph{Trade–off implies budget.}
Fix any $\gamma>0$ and let
\begin{align}
\theta_{\gamma}\;=\;\arg\min_{\theta}\big[\sigma^2(\theta)+\gamma E(\theta)\big]\ \ \text{s.t.}\ \ \mu(\theta)=\mu^{*},
\quad
E^{*}:=E(\theta_{\gamma}).
\end{align}
At $\theta_{\gamma}$ we have $\partial_{\theta}\sigma^2+\gamma\,\partial_{\theta}E+\alpha^{*}\partial_{\theta}\mu=0$ and $E(\theta_{\gamma})=E^{*}$. 
Therefore $\theta_{\gamma}$ satisfies the KKT system of the budgeted problem with \emph{the same} stationarity and complementarity, and with multiplier $\lambda^{*}=\gamma>0$.
Thus the constrained optimum with budget $E^{*}$ coincides with the trade–off optimum, and the budget is active.

\paragraph{Budget implies trade–off.}
Conversely, suppose the constrained problem with budget $E^{*}$ attains $(n^{*},p^{*},q^{*})$ with $\lambda^{*}>0$. Setting $\gamma:=\lambda^{*}$ gives
\begin{align}
\partial_{\theta}\sigma^2+\gamma\,\partial_{\theta}E+\alpha^{*}\partial_{\theta}\mu=0,
\end{align}
so $(n^{*},p^{*},q^{*})$ also satisfies the trade–off stationarity and hence is a trade–off optimum for that $\gamma$. If the budget is slack ($\lambda^{*}=0$), the energy term drops out and no $\gamma>0$ reproduces that solution.

\paragraph{Why we assume the budget is active.}
Holding $\mu=\mu^{*}$ fixed and decreasing $\sigma^{2}$ requires increasing either $n$ or $p$ and $E$ increases along any such direction. Let $\theta_{\gamma}$ be the trade–off optimiser of $\T$ for some $\gamma>0$ and set $E^{*}:=E(\theta_{\gamma})$. If the budget were slack at $\theta_{\gamma}$ ($E(\theta_{\gamma})<E^{*}$, hence $\lambda^{*}=0$), one could move a small step that lowers $\sigma^{2}$ while keeping $\mu$ fixed and still remain feasible ($E<E^{*}$), contradicting optimality of $\theta_{\gamma}$ for $\T$. Therefore the budget is \emph{active} at the shared solution ($E(\theta_{\gamma})=E^{*}$ and $\lambda^{*}>0$).

In the data-driven case we infer a finite $E^{*}$ from observed $(\mu,\sigma^{2})$; since $\sigma^{2}>0$ and $E^{*}<\infty$, the same argument applies: if $E(\theta^{*})<E^{*}$, $\sigma^{2}$ could be reduced further at fixed $\mu$ without violating the constraint, so the optimum must lie on $E=E^{*}$. Degenerate limits ($p\!\to\!1$ or $n\!\to\!\infty$) would imply $\sigma^{2}\!\to\!0$ and $E\!\to\!\infty$ and are not supported by the finite $(\mu,\sigma^{2},E^{*})$ observed, hence they do not arise in our fits.

\paragraph{Conclusion.}
Evaluated at the same optimiser with an \emph{active} budget,
\begin{align}
\boxed{\ \lambda^{*}=\gamma>0\ },
\end{align}
so the energy that minimises $\T$ equals the energy on the binding budget in $\L$. 
Equivalence fails only for slack budgets, where $\lambda^{*}=0$ and $E^{*}$ does not shape the solution.

\textbf{Minimal (active) energy boundary}
The condition $\lambda^*=\gamma>0$ implies the energy constraint is active (KKT complementarity),
and the optimisers of both the tradeoff and budget formulations lie on the \emph{minimal feasible energy boundary} $E=E^*$.
This conclusion uses only the stationarity and KKT conditions and does not depend on any other result.

To illustrate the implications concretely, in the next sections (\hyperref[key_derivations]{Appendix - Key derivations}) we provide worked examples using the calcium pump+turnover pairing to show that the minimal $E^*(\mu,\sigma^2;\beta_p)$ at fixed $(\mu,\sigma^2)$ that solves $\T$ is equivalent to the energy constraint in $\L$.

\subsection{Key derivations}
\label{key_derivations}
The solutions for \(n^*\), \(p^*\), \(q^*\), and \(\sigma^*\) depend on the choice of the energy function. In this section, we derive these solutions under a biologically plausible trade-off between synapse performance and energy cost.

\subsubsection{Turnover and calcium pump cost}
If the cost function \(\T\) includes both turnover and calcium pump costs, we can express it as:
\begin{align}
\T &= \sigma^2 + \gamma E + \text{const},
\end{align}
where \(\sigma^2\) represents synaptic variance, \(\gamma\) is a trade-off parameter, and \(E\) is the energy cost. 
Normalising by \((\mu^*)^2\):
\begin{align}
\T &\propto \frac{\sigma^2}{(\mu^*)^2} + \frac{\gamma}{(\mu^*)^2}E.
\end{align}

Expressing \(\T\) in terms of quantal parameters:
\begin{align}
\T &= \frac{1}{(\mu^*)^2} np(1-p)q^2 + \frac{\gamma}{(\mu^*)^2}E,
\end{align}
where \(b = \frac{p}{1-p}\). Substituting \(b\) and simplifying:
\begin{align}
\T &\propto \frac{1}{nb} + \frac{\gamma}{(\mu^*)^2}\big[\beta_p b^{\frac{1}{4}} + (1-\beta_p)n\big].
\end{align}

\textbf{Optimisation of \(\T\).}
To find the optimal values for \(n\) and \(b\), we take derivatives of \(\T\) with respect to \(n\) and \(b\) and set them to zero.

\textbf{(i) Derivative with respect to \(n\):}
\begin{align}
\frac{d\T}{dn} &= -\frac{1}{n^2b} + \frac{\gamma}{(\mu^*)^2}(1-\beta_p) = 0, \\
-k^2 + \frac{\gamma}{(\mu^*)^2}(1-\beta_p)n &= 0.
\end{align}

\textbf{(ii) Derivative with respect to \(b\):}
\begin{align}
\frac{d\T}{db} &= -\frac{1}{nb^2} + \frac{1}{4}\frac{\gamma}{(\mu^*)^2}\beta_p b^{-\frac{3}{4}} = 0, \\
-k^2 + \frac{1}{4}\frac{\gamma}{(\mu^*)^2}\beta_p b^{\frac{1}{4}} = 0.
\end{align}

\textbf{Solving for \(\gamma\).}
Multiplying the result of (ii) by $4$, adding (i), and recognising $E$:
\begin{align}
-k^2 + \frac{\gamma}{(\mu^*)^2}E - 4k^2 &= 0, \\
\frac{\gamma}{(\mu^*)^2} &= 5\frac{k^2}{E}, \\
\gamma &= 5\frac{\sigma^2}{E}.
\label{gamma_turnover}
\end{align}
Thus, increasing \(\gamma\) increases \(\sigma^2\) while reducing energy use, reflecting the trade-off between reliability and energy efficiency.

Using this in (ii) gives:
\begin{align}
b^{\frac{1}{4}} &= \frac{4}{5\beta_p}E, \qquad
b = \left(\frac{4}{5\beta_p}\right)^4E^4.
\end{align}

\textbf{Solution for \(n^*\):}
Returning to (i):
\begin{align}
n^* = \frac{1}{5}\frac{E}{1-\beta_p}.
\end{align}

\textbf{Combined Solution for \(k^{-2}\):}
\begin{align}
k^{-2} = n^*b^* = \frac{4^4}{5^5}\frac{(E^*)^5}{\beta_p^4(1-\beta_p)}.
\end{align}

\textbf{Energy Constraints and \(\gamma\):}
Solving for \(E^*\) in terms of \(\gamma\):
\begin{align}
E^{6} = \frac{5^6}{4^4}\beta_p^4(1-\beta_p)\frac{(\mu^*)^2}{\gamma}.
\label{gamma_e_dependence}
\end{align}
This shows that \(\gamma\) sets the minimal energy, $E$.

\paragraph{Derivation for $\frac{\beta_p}{1-\beta_p}$.}
\label{beta_i}
From $b^*$ and $n^*$:
\begin{align}
(b^*)^{\frac{1}{4}} &= \frac{4}{5\beta_p}E^* = \frac{4(1-\beta_p)}{\beta_p}n^*, \\
\frac{(b^*)^{\frac{1}{4}}}{n^*} &= 4\frac{1-\beta_p}{\beta_p}.
\end{align}
This demonstrates that $\beta_p$ alone sets the ratio between $b^*$ and $n$.

\subsubsection{Lagrangian formulation: Turnover and calcium pump cost}
\label{lagrange_formulation}
To incorporate energy constraints explicitly, we consider the Lagrangian:
\begin{align}
\L = \sigma^2 + \lambda (E - E^*) + \alpha(\mu - \mu^*).
\end{align}
Normalising by $(\mu^*)^2$ and substituting $E$:
\begin{align}
\L(n,b) &\propto \frac{1}{nb} + \frac{\lambda}{(\mu^*)^2}\big[\beta_p b^{\frac{1}{4}} + (1-\beta_p)n - E^*\big] + \text{const}.
\end{align}
We will show the active boundary $E^*$ is equivalent to the minimal energy $E(\mu^*,\sigma^{2*})$ that solves $\T$ where $\gamma=\lambda^*>0$.  
% In addition, we show that significantly for the calcium pump+turnover pairing that the minimal energy boundary
% yields
% $\beta$-invariant predictions for $\sigma$ in our two-stage energy cost selection procedure in \hyperref[energy_cost_selection]{Results - Energy cost selection}.

\textbf{Derivatives of the Lagrangian.}

\textbf{(i) Derivative with respect to \(n\):}
\begin{align}
\frac{d\L}{dn} = -\frac{1}{n^2b} + \frac{\lambda}{(\mu^*)^2}(1-\beta_p) = 0,
\qquad
-k^2 + \frac{\lambda}{(\mu^*)^2}(1-\beta_p)n = 0.
\end{align}

\textbf{(ii) Derivative with respect to \(b\):}
\begin{align}
\frac{d\L}{db} = -\frac{1}{nb^2} + \frac{1}{4}\frac{\lambda}{(\mu^*)^2}\beta_p b^{-\frac{3}{4}} = 0,
\qquad
-k^2 + \frac{1}{4}\frac{\lambda}{(\mu^*)^2}\beta_p b^{\frac{1}{4}} = 0.
\end{align}

\textbf{Solving for \(\lambda\):}
Multiplying (ii) by $4$, adding (i) and recognising E
\begin{align}
\frac{\lambda}{(\mu^*)^2} = \frac{5k^2}{E^*}, \qquad
\lambda = \frac{5}{E^*}\sigma^2 
\end{align}
Recalling the equation for $\gamma$ (Eq.~\eqref{gamma_turnover}), giving $\lambda^*=\gamma$. 

\paragraph{Equivalence with the trade-off objective.}
At $\mu=\mu^*$, if $\lambda^* = \gamma$, the gradients of $\L$ and $\T$  match:
\begin{align}
\nabla \L = \nabla \sigma^2 + \lambda^*\nabla E = 
\nabla \sigma^2 + \gamma \nabla E = \nabla \T,
\end{align}
showing that $\T$ and $\L$ yield the same optimum $(n^*,b^*,\sigma^{*})$. This means that the energy minimised through $\T$, is equal to the energy constraint in $\L$, $E(n^*,b^*)=E^*$.

\paragraph{The energy budget implicitly tunes $\gamma$.}
\label{gamma_e_tunes}
Furthermore, the equivalence between the trade-off and Lagrangian formulations can be viewed in terms of how 
the energy budget $E^*$ determines the implict effective trade-off weight $\gamma$. 
Because the optimisers of $\L$ under the constraint $E=E^*$ coincide with those of $\T$, 
the energy $E$ obtained by solving the trade-off problem must equal the budget $E^*$. 
Substituting $E=E^*$ into Eq.~\eqref{gamma_e_dependence} gives:
\begin{align}
\gamma(\mu^{*},E^{*}) = 
\frac{5^{6}}{4^{4}}\beta_p^{4}(1-\beta_p)\,\frac{(\mu^{*})^{2}}{(E^{*})^{6}}.
\end{align}
Thus, smaller budgets $E^{*}$ correspond to larger $\gamma$, weighting energy cost more heavily and 
shifting the optimum towards higher synaptic variance (lower reliability).

\subsection{Summary of results}
\label{summary_section}

Finally, we provide the derived expressions of $n^*$, $b^*$ and $\sigma^*$ for varied choices of the energy cost function $E$, 
where the calcium pump cost is paired with each of the other candidate costs in turn. 
Note that only the calcium pump+turnover and calcium pump+trafficking combinations admit fully analytical solutions; 
the calcium pump+actin and calcium pump+membrane pairings depend explicitly on $q$ and are therefore not algebraically soluble without further assumptions. 
Furthermore, only the calcium pump+turnover combination yields a simple power-law relation between $k^{-2}$ and $E^*$:
\begin{align}
k^{-2} \propto (E^*)^{5},
\qquad
E^* \propto (\sigma^{*})^{-2/5}.
\end{align}
\small
\[
\begin{array}{c|c c c }
\label{summary_table}
    & n^{*} & b^{*} & \sigma^{*-2}  \\ \hline
    \text{Turn} & 
        \frac{E^{*}}{5(1-\beta)} & 
        \left(\frac{4}{5\beta}\right)^{4}(E^{*})^{4} & 
        \frac{4^{4}}{5^{5}}\frac{(E^{*})^{5}}{\beta^{4}(1-\beta)(\mu^{*})^{2}} \\[4pt]
    \text{Actin} & 
        \frac{E^{*}}{(1-\beta)q^{1/3}\mu^{*}}
        \left(\frac{3q}{11\mu^{*} q+4{\sigma^{*}}^{2}}\right) &
        \frac{(E^{*})^{4}}{\beta^{4} (\mu^{*})^{4}}
        \left(1- \frac{3q}{11\mu^{*} q+4{\sigma^{*}}^{2}}\right)^{4} &
        \frac{(E^{*})^{5}\mu^{*}}{(1-\beta)\beta^{4}q^{1/3}}
        \left(\frac{3q}{11q \mu^{*}+4{\sigma^{*}}^{2}}\right)
        \left(1- \frac{3q}{11q\mu^{*}+4{\sigma^{*}}^{2}}\right)^{4} \\[4pt]
    \text{Memb} &  
        \frac{E^{*}}{(1-\beta)q^{2/3}\mu^{*}}
        \left(\frac{3q}{11\mu^{*} q+4{\sigma^{*}}^{2}}\right) &
        \frac{(E^{*})^{4}}{\beta^{4} (\mu^{*})^{4}}
        \left(1- \frac{3q}{11\mu^{*} q+4{\sigma^{*}}^{2}}\right)^{4} &
        \frac{(E^{*})^{5}\mu^{*}}{(1-\beta)\beta^{4}q^{2/3}}
        \left(\frac{3q}{11q \mu^{*}+4{\sigma^{*}}^{2}}\right)
        \left(1- \frac{3q}{11q\mu^{*}+4{\sigma^{*}}^{2}}\right)^{4} \\[4pt]
    \text{Traff} & 
        \frac{E^{*}}{5(1-\beta)}\!\left[\!\left(\frac{4}{5\beta}E^{*}\right)^{-4}\! +1\right] &
        \left(\frac{4}{5\beta}\right)^{4}(E^{*})^{4} &
        \frac{1}{(\mu^{*})^{2}}\!\left[\frac{E^{*}}{5(1-\beta)} 
        + \frac{(E^{*})^{5}}{5(1-\beta)}\!\left(\frac{4}{5\beta}\right)^{4}\!\right]
\end{array}
\]
\normalsize
\\
\\
\\
\noindent\textit{Equivalence and active bound:}
Because the trade-off and budget formulations share identical first-order conditions at the optimum, and the Lagrange multiplier matches the trade-off multiplier,
\begin{align}
\lambda^* = \gamma > 0,
\end{align}
the budget $E^*$ implied by the trade-off formulation always makes the energy constraint \emph{active} 
(see \hyperref[equivalence_general]{Appendix -- Equivalence of trade–off and energy–budget formulations}). 
Hence, all solutions reported in Table~\ref{summary_table} lie on the minimal feasible energy boundary ($E=E^*$) and apply 
equally under either formulation.

\subsection{Monotonicity, separability, and log–log linearity}
Among the pairings we consider, \textbf{only calcium pump+turnover} yields a strictly monotone, 
power-law scaling between synaptic reliability and the \emph{minimum feasible energy} at fixed mean $\mu^*$. 
With $b=\frac{p}{1-p}$ and $k^{-2}=nb=\mu^{2}/\sigma^{2}$, the active-boundary budget is
\begin{align}
E^*(\mu,\sigma^2;\beta_p)=\kappa_{\mathrm{turn}}(\beta_p)\,\Big(\frac{\mu^2}{\sigma^2}\Big)^{1/5}
\quad\Longleftrightarrow\quad
\log \sigma^{-2} = \log \kappa(\beta_p) + 5\log E^{*}.
\end{align}
Here, \emph{separability} means that $k^{-2}$ factorises into a constant depending only on 
$(\mu^*,\beta_p)$ and a pure power of $E^{*}$, so that the effect of $\beta_p$ is a simple 
vertical rescaling of the $\log \sigma^{-2}$–$\log E^{*}$ maintaining $slope=5$. This constant slope on the log-log scale could be tested empirically. 
Crucially, a multiplicative change $E^{*}\mapsto hE^{*}$ produces a constant vertical shift:
\begin{align}
\log \sigma^{-2}(hE^{*}) = \log \sigma^{-2}(E^{*}) + 5\log h.
\end{align}
Thus, the precision gain from scaling energy is \emph{predictable and uniform}. Every 
doubling ($h=2$) of $E^{*}$ yields a $2^{5}$-fold increase in $\sigma^{-2}$, regardless of the 
starting budget. This log–log linearity makes energy allocation globally consistent. 
Synapses at different baseline budgets respond by the same factor to a given multiplicative change in energy.

By contrast, calcium pump+trafficking and the $q$-dependent pairings (calcium pump+actin/membrane) involve 
additional nonlinear dependencies on $(\beta_p, b, q)$ that distort the mapping between 
energy and reliability. In those cases, $\beta_p$ not only rescales the solution but also 
changes the \emph{functional shape} of the curve, so the reliability gain from a fixed 
energy increase depends on the baseline $E^{*}$. This makes budget allocation less 
uniform across synapses. The unique power-law scaling under calcium pump+turnover therefore 
offers both analytic simplicity and a biologically appealing principle: a consistent 
and interpretable link between energy investment and achievable precision.

Next, we show---importantly for our two-stage energy-cost selection procedure in \hyperref[methods_energy_cost]{Methods -- Energy cost selection and rationale}---that for the calcium pump+turnover pairing the minimal energy boundary
yields $\beta_p$-invariant predictions for $\sigma^2$. This is a direct consequence of separability.

\subsection{$\beta_p$-invariance for calcium pump+turnover}
\label{beta_invariance}

\paragraph{Setup.}
For the calcium pump+turnover pairing,
\begin{align}
E(n,b;\beta_p)=\beta_p\,b^{1/4}+(1-\beta_p)\,n,\qquad
k^{-2}=nb=\frac{\mu^2}{\sigma^2},\quad b=\frac{p}{1-p}.
\end{align}
In \hyperref[methods_energy_cost]{Methods -- Energy cost selection and rationale} we first minimise $E$ subject to fixed $(\mu,\sigma^2)$ (equivalently fixed $k^{-2}=nb$) to obtain
the {active minimal boundary} $E^*(\mu,\sigma^2;\beta_p)$.

\paragraph{Minimal energy at fixed $(\mu,\sigma^2)$.}
Impose $nb=k^{-2}$, i.e.\ $n=k^{-2}/b$, and minimise
$E(b)=\beta_p\,b^{1/4}+(1-\beta_p)\,k^{-2}/b$ over $b>0$.
Setting $dE/db=0$ gives
\begin{align}
\frac{\beta_p}{4}\,b^{-3/4}-(1-\beta_p)\,k^{-2}\,b^{-2}=0
\ \Longrightarrow\
b^{5/4}=\frac{4(1-\beta_p)}{\beta_p}\,k^{-2}.
\end{align}
Hence
\begin{align}
b^*=\Big[\tfrac{4(1-\beta_p)}{\beta_p}\,k^{-2}\Big]^{4/5},\qquad
n^*=\frac{k^{-2}}{b^*}=\Big[\tfrac{\beta_p}{4(1-\beta_p)}\Big]^{4/5}\,k^{-6/5}.
\end{align}
Substitute into $E$:
\begin{align}
E^*(\mu,\sigma^2;\beta_p)
= \underbrace{\Big[\tfrac{4}{5}\,\beta_p^{-1}\big(5(1-\beta_p)\big)^{4/5}\Big]}_{\displaystyle \kappa(\beta_p)}
\Big(\frac{\mu^2}{\sigma^2}\Big)^{1/5}.
\end{align}
Thus the active boundary is \emph{separable}:
\begin{align}
E^*=\kappa(\beta_p)\,\Big(\frac{\mu^2}{\sigma^2}\Big)^{1/5}
\quad\Longleftrightarrow\quad
\sigma^{-2}=\kappa(\beta_p)^{-5}\,\frac{(E^*)^{5}}{\mu^2}.
\end{align}

\paragraph{Two-stage pipeline and $\beta$-invariance of $\sigma$.}
Our energy-cost fitting used two stages (see \hyperref[methods_energy_cost]{Methods -- Energy cost selection and rationale}): (i) minimise $E$ at $(\mu_d,\sigma_d^2)$ to obtain $E_d^*$,
then (ii) minimise $\sigma^2$ at $\mu=\mu_d$ with $E=E_d^*$.
From the boundary above, step (i) gives
\begin{align}
E_d^*=\kappa(\beta_p)\Big(\frac{\mu_d^2}{\sigma_d^2}\Big)^{1/5}.
\end{align}
Step (ii) inverts the same mapping:
\begin{align}
\sigma^{2*}
=\kappa(\beta_p)^5\,\frac{\mu_d^2}{(E_d^*)^{5}}
=\kappa(\beta_p)^5\,\frac{\mu_d^2}{\big[\kappa(\beta_p)(\mu_d^2/\sigma_d^2)^{1/5}\big]^5}
=\sigma_d^2.
\end{align}
Hence the factors $\kappa(\beta_p)$ cancel exactly and the predicted $\sigma^{2*}$ is
\emph{independent of $\beta_p$} under this two-stage procedure.

This explains the robustness of $\sigma^{2}$ predictions for varied $\beta_p$ in Methods---Figure \ref{fig_nrmse} (Calcium pump+turnover).

\paragraph{Why this is specific to calcium pump+turnover.}
For calcium pump+trafficking, \(k^{-2}\) remains a monotone function of \(E^*\) but is not
separable (it contains linear \(+\) quintic terms in \(E^*\)), so the inversion used
above does not reduce to a simple cancellation of \(\kappa(\beta_p)\).
For the \(q\)-dependent pairings (calcium pump+actin/membrane), the minimal-\(E\) and
minimal-\(\sigma^2\) problems can select different \(q\) unless extra conditions hold;
the mapping depends on \(q\) and \(\beta_p\) in a coupled way, so \(\beta_p\)-invariance of
\(\sigma\) (or \(\sigma^2\)) is not guaranteed.
This is the core reason calcium pump+turnover yields both (i) a clean active boundary and
(ii) \(\beta_p\)-invariant \(\sigma\) under our two-stage pipeline.

\subsection{Convexity conditions of the tradeoff}
\label{app:convexity}
\paragraph{Motivation for convexity analysis.}
First–order conditions provide {candidate} optima for \((n,b)\) in terms of \(\mu^*\), \(\lambda\) (or \(\gamma\)), and—under budgets—\(E^*\). However, stationarity alone does not ensure these candidates are \emph{unique} minimisers: if the objective is not (jointly) convex, the landscape may contain multiple local minima and saddle points. This matters in two ways. \emph{Numerically}, gradient–based methods (and our visual descent trajectories) converge to a nearby local minimum; without convexity, solutions can depend on initial conditions. \emph{Biologically}, plasticity must traverse the same landscape; nonconvex regions could slow or trap learning as budgets change. For these reasons we characterise convexity: (i) on the full \((n,b)\) plane via the Hessian test (yielding, for calcium pump+turnover, the explicit condition \(A\,\beta_p\,n\,b^{5/4}\le 8\) with \(A=\gamma/(\mu^*)^2\)), which certifies \emph{local} convexity at the stationary point; and (ii) along the admissible budget curve \(E(n,b)=E^*\), where the objective reduces to a one–dimensional problem that we show is \emph{strictly} convex on the plotted intervals, ensuring a \emph{unique} constrained minimum. This analysis justifies both the uniqueness claims we make under fixed budgets and our use of a constrained trust–region optimiser that descends \emph{on} the budget curve where convexity holds.

We study convexity of the objective on the fixed-mean ($\mu=\mu^*$) surface,
\begin{align}
\T(n,p,q)\;=\;\sigma^2 + \gamma\,E + \text{const},
\qquad
\sigma^2/(\mu^*)^2=\frac{1-p}{np},
\end{align}
and use the reparametrisation \(b=\tfrac{p}{1-p}>0\) (so \(p=\tfrac{b}{1+b}\)) and \(n>0\). For the illustrative \emph{calcium pump+turnover} pairing,
\begin{align}
E(n,b)=\beta_p\,b^{1/4}+(1-\beta_p)n,
\qquad 0<\beta_p<1,
\end{align}
dividing by $\mu^*$, the normalised objective takes the form
\begin{align}
f(n,b)
~\propto~
\underbrace{\frac{1}{n\,b}}_{\text{noise}}
~+~
\frac{\gamma}{(\mu^*)^2}\underbrace{\Big(\beta_p\,b^{1/4}+(1-\beta_p)n\Big)}_{\text{energy}},
\quad n>0,~b>0.
\end{align}
Writing \(A\equiv\gamma/(\mu^*)^2>0\).
The Hessian of \(f(n,b)\) is
\begin{align}
\bm{H}(n,b)=\begin{pmatrix}
\frac{\partial ^2f}{\partial n^2} & \frac{\partial ^2f}{\partial bn}\\[4pt]
\frac{\partial ^2f}{\partial nb} & \frac{\partial ^2f}{\partial b^2}
\end{pmatrix}=
\begin{pmatrix}
\frac{2}{n^3 b} & \frac{1}{n^2 b^2}\\[4pt]
\frac{1}{n^2 b^2} & \frac{2}{n b^3}-\frac{3}{16}A\,\beta_p\,b^{-7/4}
\end{pmatrix}.
\end{align}
Formally, for convexity the sufficient condition is that principal minors be $\geq0$. For our $2\times2$ Hessian, the first principal minor is $H_{11}$. The second is $\det(\bm{H})$. Since \(H_{11}>0\) on the domain of $b$ and $n$, joint convexity is equivalent to the determinant being nonnegative:
\begin{align}
\det H(n,b)=H_{11}H_{22}-H_{21}H_{12}\;\ge\;0
\quad\Longleftrightarrow\quad
A\,\beta_p\,n\,b^{5/4}\;\le\;8.
\label{eq:convexity_condition}
\end{align}

The joint--convexity condition for the calcium pump+turnover case is
\(A\,\beta_p\,n\,b^{5/4}\le 8\) with \(A=\gamma/(\mu^*)^2>0\) (cf.\ \eqref{eq:convexity_condition}).
Because the left-hand side can be made arbitrarily large by increasing \(n\) and/or \(b\),
the condition cannot hold over the entire domain \(n>0,b>0\).
Hence, the objective is \emph{not globally convex}.

\paragraph{Numerical check.}
To document convexity in any plotted panel, one may report the maximal value of our convexity condition,
\begin{align}
\mathcal{C}(n,b,A)=A\,\beta_p\,n\,b^{5/4}\leq 8
\end{align}
by substituting numerical values for $A=\gamma/(\mu^*)^2$, $n$ and $b$ over the displayed domain and verify it is \(\le 8\).

\paragraph{Energy budget dependent condition.}
For panels using the calcium pump+turnover pairing, the joint–convexity condition derived in Eq.~(\ref{eq:convexity_condition}) can be written in terms of the \emph{implied} optimal energy $E^*$ as
\begin{align}
\mathcal{C}(n,b;E^*) \;=\; A(E^*)\,\beta_p\,n\,b^{5/4}
\;\le\; 8,
\qquad
A(E^*) \;=\; \frac{\gamma}{(\mu^*)^2}
\;=\; \frac{(5^6/4^4)\,\beta_p^4(1-\beta_p)}{(E^*)^6}.
\end{align}
Thus the single quantity to check is
\begin{align}
{\;\mathcal{C}(n,b;E^*) \;=\; 
\frac{(5^6/4^4)\,\beta_p^5(1-\beta_p)}{(E^*)^6}\; n\,b^{5/4}\;\leq 8}
\end{align}
and convexity on the displayed domain is guaranteed whenever $\mathcal{C}\le 8$ throughout.

Equivalently, one can report the minimum energy threshold that would still guarantee joint convexity over the plotted ranges:
\begin{align}
E^* \;\ge\; 
\left[\frac{(5^6/4^4)\,\beta_p^5(1-\beta_p)}{8}\; n\,b^{5/4}\right]^{\!1/6}.
\end{align}
Therefore, if the upper bounds of $n$ and $b$ grow, the energy budget needs to be increased to meet the convexity condition across the sets of $n$, $b$. Conversely, smaller energy budgets shrink the convex set.

This shows the intended behaviour of the energy budgets: Increasing $E^*$ expands the convex region (so larger $(n,b)$ ranges remain convex), while small $E^*$ may still yield convexity for correspondingly small $(n,b)$ typical of that budget. Practically, this justifies using local descent initialised within the empirically relevant $(n,b)$ ranges for each budget. A general practical measure that promotes convergences to the global minimum, and improves optimiser stability, is to initialise the dynamics at small $n$ and $b$ especially when budgets are tight.

\paragraph{Numerical check (Fig.~\ref{fig_loss}).}
\label{fig_loss_check}
For the panel parameters $\gamma=0.25$, $\mu^*=0.5$, and $\beta_p=0.7$, the trade–off weight is
\begin{align}
A \;=\; \frac{\gamma}{(\mu^*)^2} \;=\; \frac{0.25}{0.5^2} \;=\; 1.
\end{align}
On the fixed–mean surface, joint convexity of $f(n,b)$ (calcium pump+turnover) requires
\begin{align}
\mathcal{C}(n,b)\;=\;A\,\beta_p\,n\,b^{5/4}\;\le 8,\qquad b=\frac{p}{1-p}.
\end{align}
Using the ranges shown in Fig.~\ref{fig_loss} (approximately $n\in[1,2]$, $p\in[0.5,0.9]$),
the corner value at $n_{\max}=2$, $p_{\max}=0.9$ (so $b_{\max}=0.9/0.1=9$) is
\begin{align}
\mathcal{C}_{\text{corner}} \;=\; 1 \times 0.7 \times 2 \times 9^{5/4}
\;\approx\; 19.5 \;>\; 8,
\end{align}
and at $p_{\max}=0.95$ it rises further ($\mathcal{C}\approx 45$). Thus the sufficient joint-convexity test does \emph{not} hold everywhere on the plotted rectangle—consistent with the fact that the surface need not be globally convex.

By contrast, at the \emph{optimum} the Hessian is positive definite. From the closed-form solution,
\begin{align}
E^*=\Big(\tfrac{5^6}{4^4}\,\beta_p^4(1-\beta_p)\,\tfrac{(\mu^*)^2}{\gamma}\Big)^{1/6}
\approx 1.28,\quad
n^*=\tfrac{E^*}{5(1-\beta_p)}\approx 0.85,\quad
(b^*)^{1/4}=\tfrac{4}{5\beta_p}E^*\approx 1.46,
\end{align}
which gives $b^*\approx 4.58$ and $p^*=b^*/(1+b^*)\approx 0.82$. Evaluating the convexity score at the optimum,
\begin{align}
\mathcal{C}(n^*,b^*) \;=\; A\,\beta_p\,n^*\,(b^*)^{5/4} \;=\; 4 \;<\; 8,
\end{align}
verifying local convexity in the neighbourhood of the minimum marked by the black cross.

Although the heat maps over $\T$ in Fig.~\ref{fig_loss} appear “bowl–shaped,” our Hessian test shows the surface is not \emph{globally} convex on the whole plotting rectangle: $\mathcal{C}(n,b;E^*)>8$ is attained only near the high–$n$, high–$p$ (large $b$) corner, while the basin around the minimum satisfies $\mathcal{C}\le 8$. This mismatch is expected: (i) the convexity test is local (based on $\det H$), so small saddle regions can exist far from the optimum; (ii) color quantisation and interpolation visually smooth shallow negative curvature.

\medskip

\end{appendixbox}
\end{document}